\documentclass[twocolumn,aps,prd,preprintnumbers,superscriptaddress,nofootinbib]{revtex4-1}
\usepackage{aas_macros}
\usepackage{here}
\usepackage{hyperref}
\usepackage{amsmath,amssymb,amsfonts,mathrsfs}
\usepackage[dvipdfmx]{graphicx}
\usepackage{color}
\usepackage{enumerate} 
\usepackage{bm}
\usepackage{array}
\usepackage{siunitx}
\usepackage{float}

\definecolor{mygreen}{RGB}{0,115,0} 
 
\begin{document}
\title{Impacts of overlapping gravitational-wave signals on the parameter estimation: Toward the search for cosmological backgrounds}

%
%
\author{Yoshiaki Himemoto}
\affiliation{Department of Liberal Arts and Basic Sciences, College of Industrial Technology, Nihon University, Narashino, Chiba 275-8576, Japan}
\author{Atsushi Nishizawa}
\affiliation{Research Center for the Early Universe (RESCEU), School of Science, The University of Tokyo, Tokyo 113-0033, Japan}
\author{Atsushi Taruya}
\affiliation{Center for Gravitational Physics, Yukawa Institute for Theoretical Physics, Kyoto University, Kyoto 606-8502, Japan}
\affiliation{
Kavli Institute for the Physics and Mathematics of the Universe, Todai Institutes for Advanced Study, the University of Tokyo, Kashiwa, Chiba 277-8583, Japan (Kavli IPMU, WPI)}
%
%
%
%
%
\date{\today}
\begin{abstract} 
Third-generation gravitational wave detectors, such as the Einstein Telescope and Cosmic Explorer, will detect a bunch of  gravitational-wave (GW) signals originating from the coalescence of binary neutron star (BNS) and binary black hole (BBH) systems out to the higher redshifts, $z\sim 5-10$.  There is a potential concern that some of the GW signals detected at a high statistical significance eventually overlap with each other, and the parameter estimation of such an overlapping system can differ from the one expected from a single event. Also, there are certainly overlapping systems in which one of the overlapping events has a low signal-to-noise ratio $\lesssim 4$, and is thus unable to be clearly detected. Those system will potentially be misidentified with a single GW event, and the estimated parameters of binary GWs can be biased. We estimate the occurrence rate of those overlapping events. We find that the numbers of overlapping events are $\sim 200$ per day for BNSs and a few per hour for BBHs. Then we study the statistical impacts of these overlapping GWs on a parameter estimation based on the Fisher matrix analysis. Our finding is that the overlapping signals produce neither large statistical errors nor serious systematic biases on the parameters of binary systems, unless the coalescence time and the redshifted chirp masses of the two overlapping GWs are very close to each other, i.e., $|\mathcal{M}_{z1}-\mathcal{M}_{z2}|\lesssim10^{-4} \,(10^{-1})\,M_\odot$ and $|t_{\rm c1}-t_{\rm c2}|\lesssim10^{-2}\,(10^{-1})$\,s for BNSs (BBHs). The occurrence rate of such a closely overlapping event is shown to be much smaller than one per year with the third-generation detectors. 
\end{abstract}

\preprint{YITP-21-14}
\maketitle

\section{Introduction}
\label{sec:intro}
%

Since the first direct detection, the second-generation gravitational-wave (GW) detectors, Advanced LIGO~\cite{TheLIGOScientific:2014jea} and Advanced Virgo~\cite{TheVirgo:2014hva}, have detected a number of GWs from mergers of compact binaries \cite{LIGOScientific:2018mvr, Abbott:2020niy}. The event rate is currently $\sim 1.5$ per week, but it will increase with the improvement of detector sensitivities in the future. In the 2030s, third-generation (3G) detectors such as the Einstein Telescope (ET) \cite{ET:2020} 
and the Cosmic Explorer \cite{Evans:2016mbw} will be constructed, and they will detect $\sim 10^5$ binaries at cosmological distances in a year, 
corresponding to one detection per $\sim 100\,{\rm s}$ (e.g., Ref.~\cite{Nishizawa:2016kba}). 
In such a case, there will be events in which two GW signals accidentally overlap with each other, and it would be a potential concern that the occurrence of those events affects the detection and parameter estimation of GW signals \cite{Samajdar:2021egv, Pizzati:2021gzd}. Furthermore, through their imperfect parameter estimation, overlapping events can also affect the detection of a stochastic GW background of cosmological origin (for reviews, see, e.g., Refs.~\cite{Maggiore:1999vm,2018CQGra..35p3001C}).

Toward a solid detection of cosmological GW backgrounds in the presence of astrophysical foregrounds, the subtraction of the individual astrophysical signals would be of critical importance if the amplitude of stochastic backgrounds is comparable to or smaller than that of the foreground GWs. 
The methodology to subtract the foreground GW signals was developed in Ref.~\cite{Cutler:2006} in the case of space-based detectors and was later demonstrated by Harms {\it et al}.~\cite{Harms:2008xv} using numerical simulations. 
The efficiency of foreground subtraction crucially depends on detector configurations \cite{Yagi:2011wg, Nishizawa:2011eq, Adams:2013qma}. Further, space-based detectors will observe a number of demagnified GWs using the gravitational lensing effect \cite{Seto:2009bf}, and their impact remains unexplored. 
On the other hand, the sensitivity of the 3G ground-based detectors to a cosmological GW background was studied by Refs.~\cite{Regimbau:2016ike, Sachdev:2020bkk}, in which foreground GW signals having a signal-to-noise ratio (SNR) larger than $12$ are assumed to be perfectly resolved, and hence removable. In Ref.~\cite{Sharma:2020btq}, the improvement of the  sensitivity to a stochastic GW background was investigated by simulating mock data in the presence of the  binary black hole (BBH) foreground and subtracting them using the method developed for space-based detectors, concluding that the ultimate sensitivity of the stochastic GW search is limited not by residuals left after subtractions but by a part of the astrophysical foreground that cannot be detected, i.~e.,~GW events having a small SNR below the detection threshold.
 
Note, however, that BBHs are not the only source of foreground GWs. Instead, the GWs coming from binary neutron stars (BNSs) may constitute a more serious GW foreground. Compared to a BBH, the merger rate of the BNS is much larger, and the signal duration is longer. Thus, if the two GW signals coming from the BNS overlap over time, the interference of the overlapped waveform  may severely affect the parameter estimation. This is in fact a nontrivial issue and can potentially increase the residual noise after subtraction. To estimate the sensitivity of future detectors to a cosmological GW background, a more careful study on the signal subtraction has to be examined not only for BBHs but also for BNSs in a coherent manner.

 
 In this paper, as a first step toward the optimal subtraction of the astrophysical foregrounds, we 
investigate the impacts of the overlapping BBH and BNS signals on the parameter estimation study.
Note that the overlapping GW signals were previously investigated 
in Ref.~\cite{Crowder:2004ca}, which focused on the white dwarf binaries observed by LISA. 
More recently, Refs.~\cite{Samajdar:2021egv, Pizzati:2021gzd} studied the overlapping GWs from BBHs and BNSs detected with third-generation detectors.
Based on the Bayesian framework, Refs.~\cite{Samajdar:2021egv, Pizzati:2021gzd} focused on the systematic bias in the parameter estimation for a typical pair of overlapping GW events. Here, our interest lies in the foreground noise subtraction, and we are particularly concerned with a different parameter region where all parameters of the overlapping binary systems are rather close to each other. In this respect, this paper covers the worst cases , which Refs.~\cite{Samajdar:2021egv, Pizzati:2021gzd} did not explore. Although such events do not frequently happen, the expected number of events is not entirely negligible during the year of observation, and it is thus important to clarify their impacts.

\begin{figure}[t]
\begin{center}
\includegraphics[width=7cm]{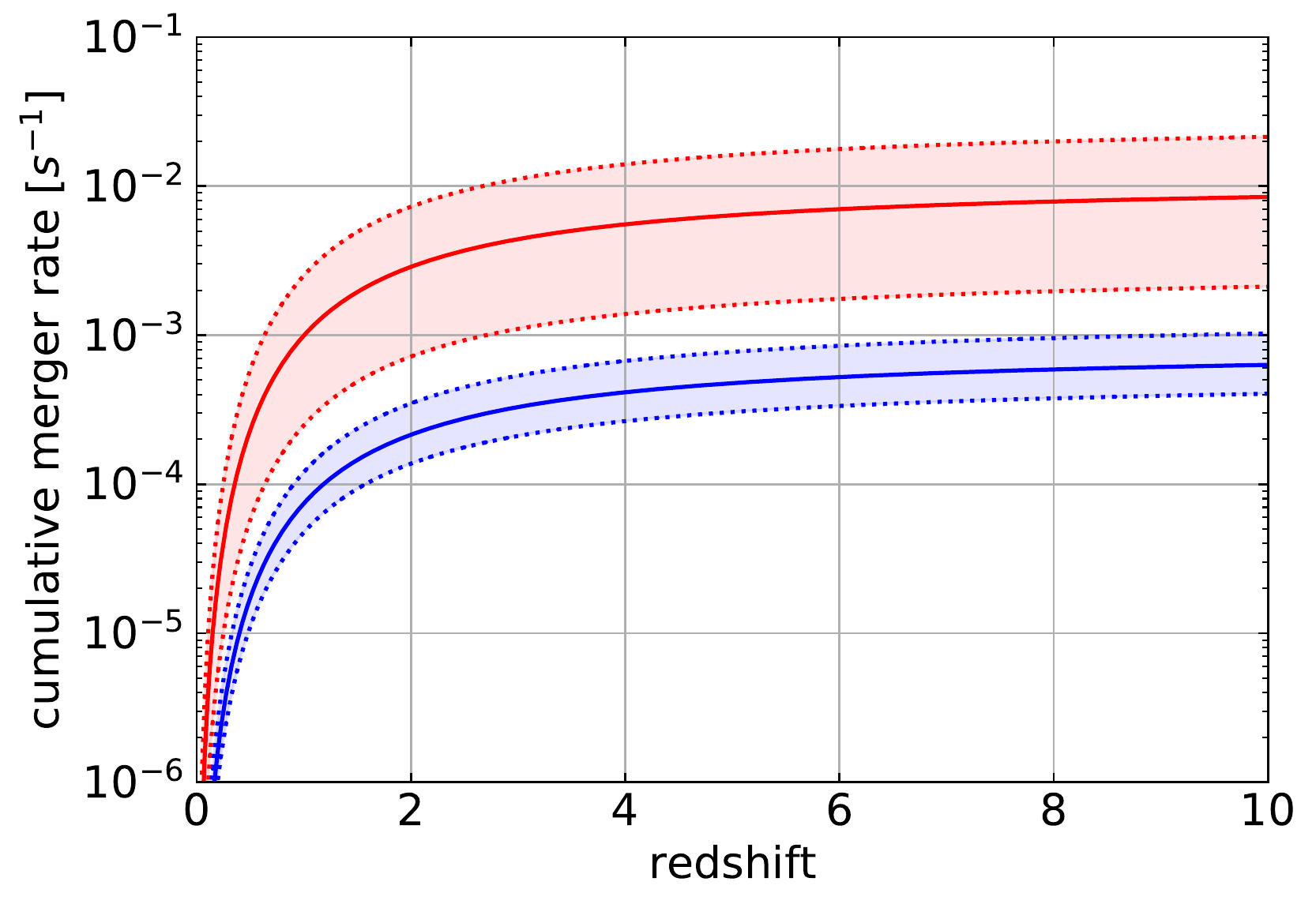}
\caption{Cumulative merger rates of BNSs (red) and BBHs (blue) as a function of redshift when one 
considers the observational uncertainty in the merger rates. The solid lines represent the results adopting the median values of the merger rates, and the shaded region indicates the 90\% credible bounds.}
\label{fig:merger-rate-z}
\end{center}
\end{figure}

 
The organization of the paper is as follows. In Sec.~\ref{sec:parameter-dist}, we generate the distributions of redshift and chirp mass for BNSs and BBHs using the Monte Carlo method and based on their realistic merger rates. Then, assuming a year of observation, we estimate the occurrence rate of the overlapping GW events in which the parameters of each binary system are close to each other. In Sec.~\ref{sec:Fisher_Matrix}, we review the formalism of the Fisher information matrix to estimate the statistical error and systematic bias in the parameter estimation. We then derive the criteria that the impacts of the overlapping waveforms on the parameter estimation are significant. In Sec.~\ref{sec:results}, the forecasted results based on the Fisher matrix formalism are presented and compared to the criteria derived in Sec.~\ref{sec:Fisher_Matrix}. Finally, Sec.~\ref{sec:conclusion} is devoted to the conclusion and a discussion. Throughout the paper, we adopt the units $c=G=1$.

\begin{figure}[b]
\begin{center}
\includegraphics[width=7cm]{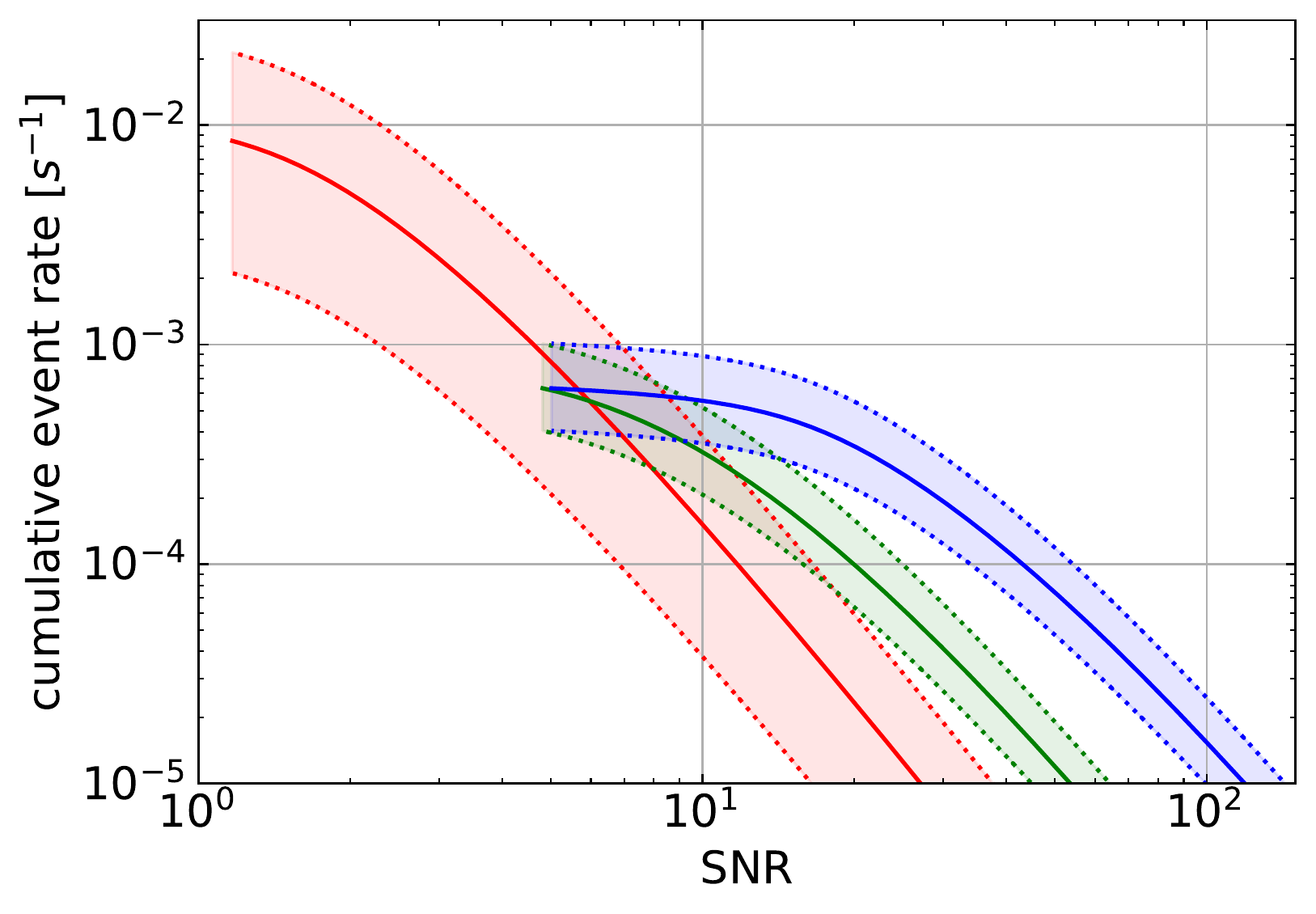}
\caption{Cumulative event rate of $1.33\,M_{\odot}$-$1.33\,M_{\odot}$ BNSs (red), $10\,M_{\odot}$-$10\,M_{\odot}$ BBHs (green), and $30\,M_{\odot}$-$30\,M_{\odot}$ BBHs (blue) as a function of SNR.}
\label{fig:merger-rate-snr}
\end{center}
\end{figure}

\section{Distribution of compact binaries}
\label{sec:parameter-dist}

\begin{figure*}[htb!]
\begin{center}
\includegraphics[width=13.5cm]{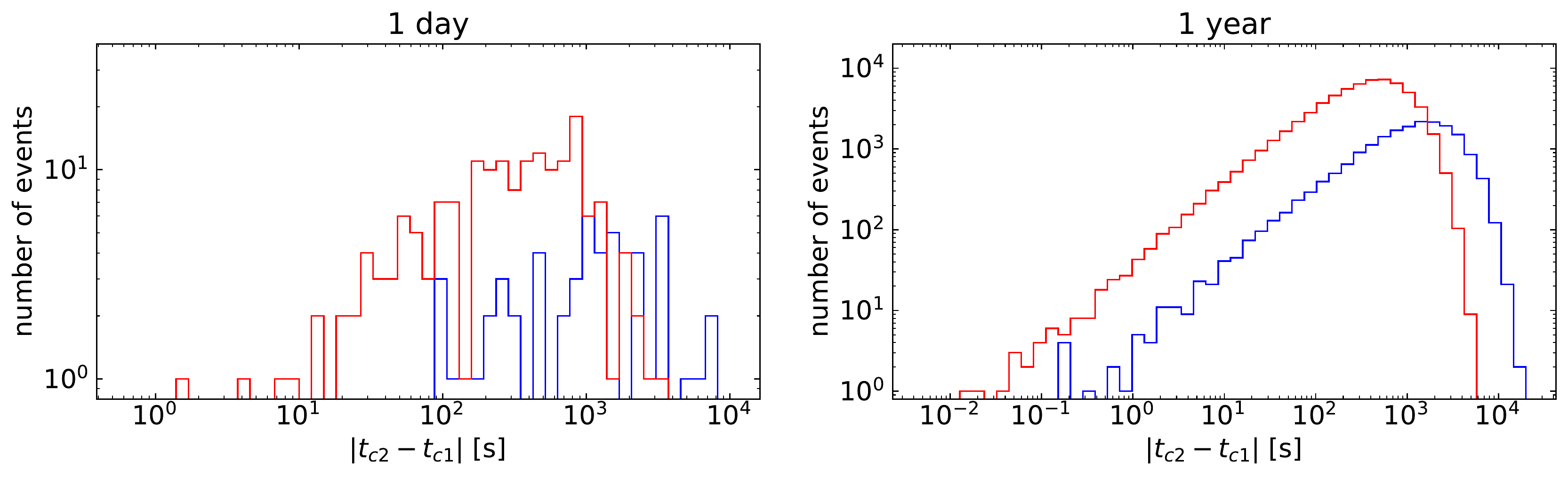}
\caption{Number-of-event distributions of the coalescence-time difference, $|t_{\rm c2}-t_{\rm c1}|$, for BNSs (red lines) and BBHs (blue lines) generated by assuming an event rate with a SNR$>4$.}
\label{fig:tc-diff-dist}
\end{center}
\end{figure*}

\begin{figure*}[htb!]
\begin{center}
\includegraphics[width=14cm]{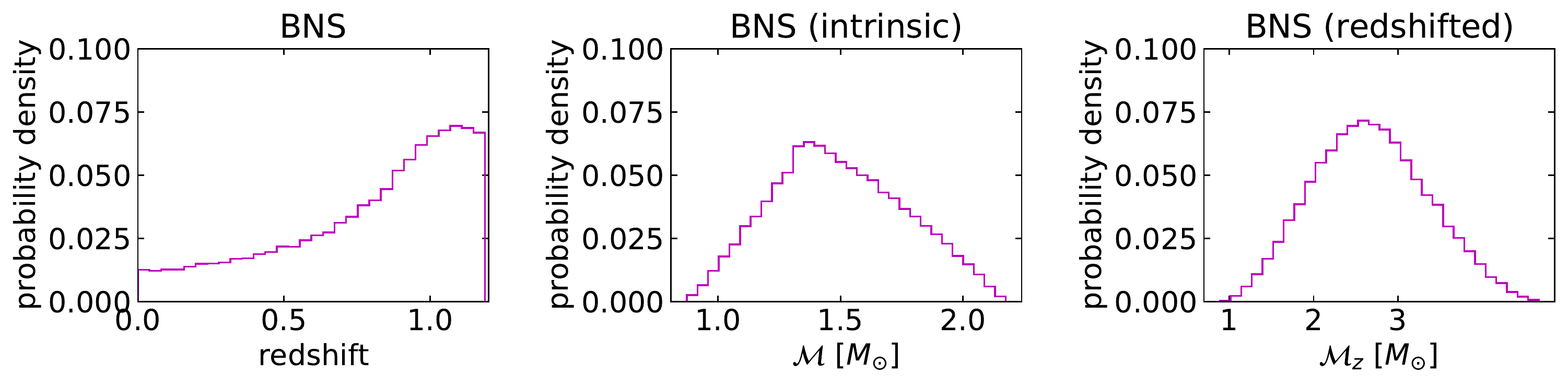}\\
\hspace{2.5mm}
\includegraphics[width=14cm]{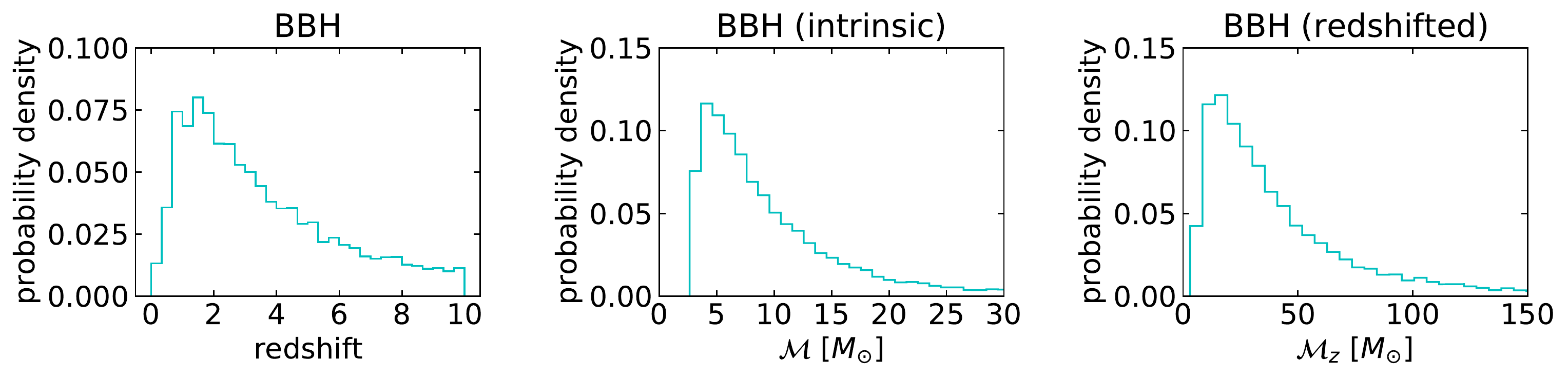}
\caption{Probability density distributions of (left panels) the redshifts, (middle and right panels) the intrinsic and redshifted chirp masses, which we respectively denote by ${\cal M}$ and ${\cal M}_z\equiv(1+z){\cal M}$, for (top row) BNSs and (bottom row) BBHs  generated by assuming an event rate with a SNR$>4$.}
\label{fig:z-mc-dist}
\end{center}
\end{figure*}
In this section, we quantify the occurrence rate of two GW signals that incidentally overlap, focusing particularly   on 3G detectors. For this purpose, 
following Refs.~\cite{Nishizawa:2016kba, Nishizawa:2016ood},
we generate source distributions with the Monte Carlo method and estimate the number and properties of the overlapping signals.
The intrinsic number of sources at redshift between $z$ and $z+dz$ per unit time in the detector frame is written as
\begin{equation}
\frac{dN}{dz dt} = \frac{4\pi \chi^2(z)}{(1+z)H(z)} \dot{n}(z) \;,
\label{eq:diff_number_dist}
\end{equation}
where $\chi(z)$ is the comoving distance to the sources, $H(z)$ is the Hubble parameter, and $\dot{n}(z)$ is the merger rate per unit comoving volume and unit proper time at redshift $z$. We use the cosmological parameters determined by the Planck Collaboration and galaxy surveys \cite{Planck2018cosmology}. Since the redshift dependence of $\dot{n}(z)$ is highly uncertain, we assume a constant merger rate, $\dot{n}(z)=\dot{n}_0$. The merger rates for BBHs and BNSs have been constrained by GW events in GWTC-1 and GWTC-2 \cite{Abbott:2020gyp}. For BBHs, we adopt $\dot{n}_0=23.9^{+14.3}_{-8.6}\,{\rm Gpc}^{-3}\,{\rm yr}^{-1}$ while assuming a broken power-law mass distribution.
According to  Ref.~\cite{Abbott:2020gyp}, the power-law index  changes below and above $39.7\,M_{\odot}$. However, the number of events above $39.7\,M_{\odot}$ is sufficiently small, and in our analysis we simply assume the single-power law of $\alpha=1.58$ and consider BH mass up to $20\,M_{\odot}$.
For BNSs, we consider a uniform mass distribution in the range of $[1\,M_{\odot}, 2.5\,M_{\odot}]$ and adopt  $\dot{n}_0=320^{+490}_{-240}\,{\rm Gpc}^{-3}\,{\rm yr}^{-1}$.  In Fig.~\ref{fig:merger-rate-z}, integrating Eq.~(\ref{eq:diff_number_dist}) over redshifts, the cumulative merger rate out to a redshift $z$ is obtained and the results for BBHs and BNSs are shown. 

To convert the source redshift of the cumulative merger rate to SNR, one needs a detector noise curve and binary masses. In this paper, we consider a single equilateral-triangle configuration of the Einstein Telescope and adopt the fitting formula of the noise power spectrum (ET-D), given in Appendix B of Ref.~\cite{Nishizawa:2019rra}. 
Using this noise spectrum, the SNRs are estimated for $1.33\,M_{\odot}$-$1.33\,M_{\odot}$ BNSs, and  $10\,M_{\odot}$-$10\,M_{\odot}$ and $30\,M_{\odot}$-$30\,M_{\odot}$ BBHs \footnote{For the SNR computation, we use the sky-averaged aligned-spinning inspiral (TaylorF2) waveform while setting the spins to zero. The minimum frequency is at $1\,{\rm Hz}$ and the maximum frequency is at twice the frequency of the innermost stable circular orbit, -namely, the innermost stable circular orbit (ISCO) frequency defined in Sec.~\ref{subsubsec:statistical errors in overlapping signals}.}.
In  Fig.~\ref{fig:merger-rate-snr}, the cumulative merger rate as a function of the SNR is shown. There are the lower cutoffs of a SNR $\sim 1$ for BNSs and a SNR\,$\sim 5$ for BBHs. This is because we do not consider sources at $z>10$, where the existence of compact binaries depends on the formation scenario and is highly uncertain. Because of the difference of intrinsic merger rates, detectable events coming from the BNSs and BBHs become different, and the GWs from BNSs dominate the events, while the signals from BBHs are more sporadic.

\begin{figure*}[htb!]
\begin{center}
\includegraphics[width=13cm]{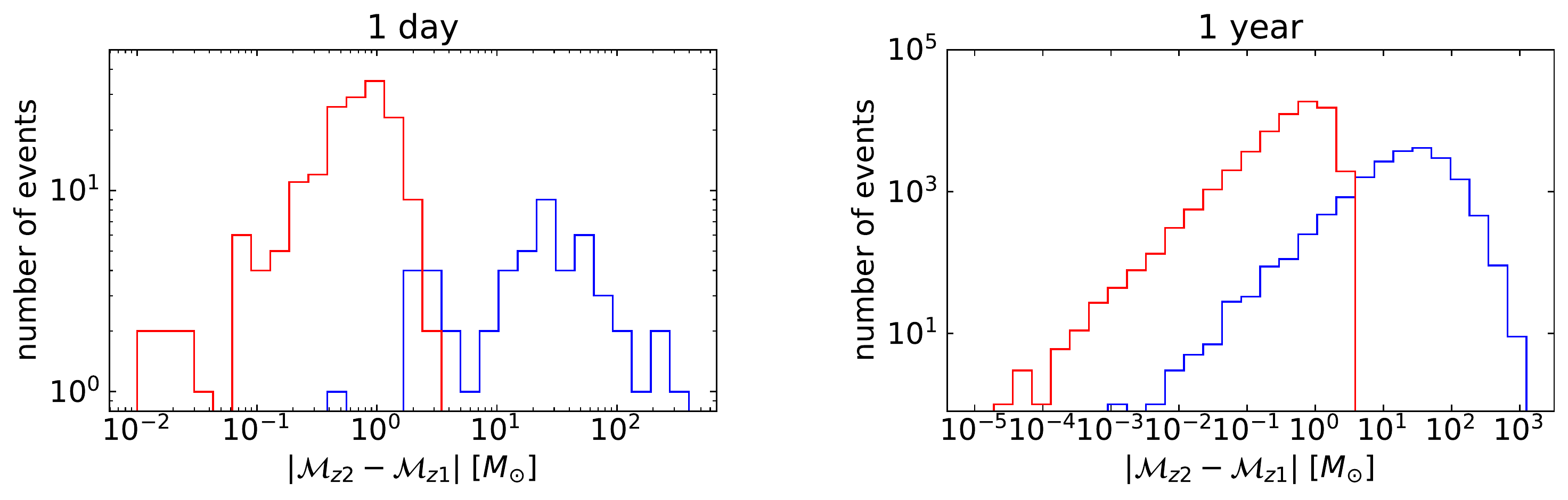}
\caption{Number-of-event distributions of the redshifted chirp-mass difference, $|{\cal M}_{z 2}-{\cal M}_{z1}|$, for BNSs (red lines) and BBHs (blue lines) generated by assuming an event rate with a SNR$>4$.}
\label{fig:mc-diff-dist}
\end{center}
\end{figure*}

\begin{figure*}[htb!]
\begin{center}
\includegraphics[width=13cm]{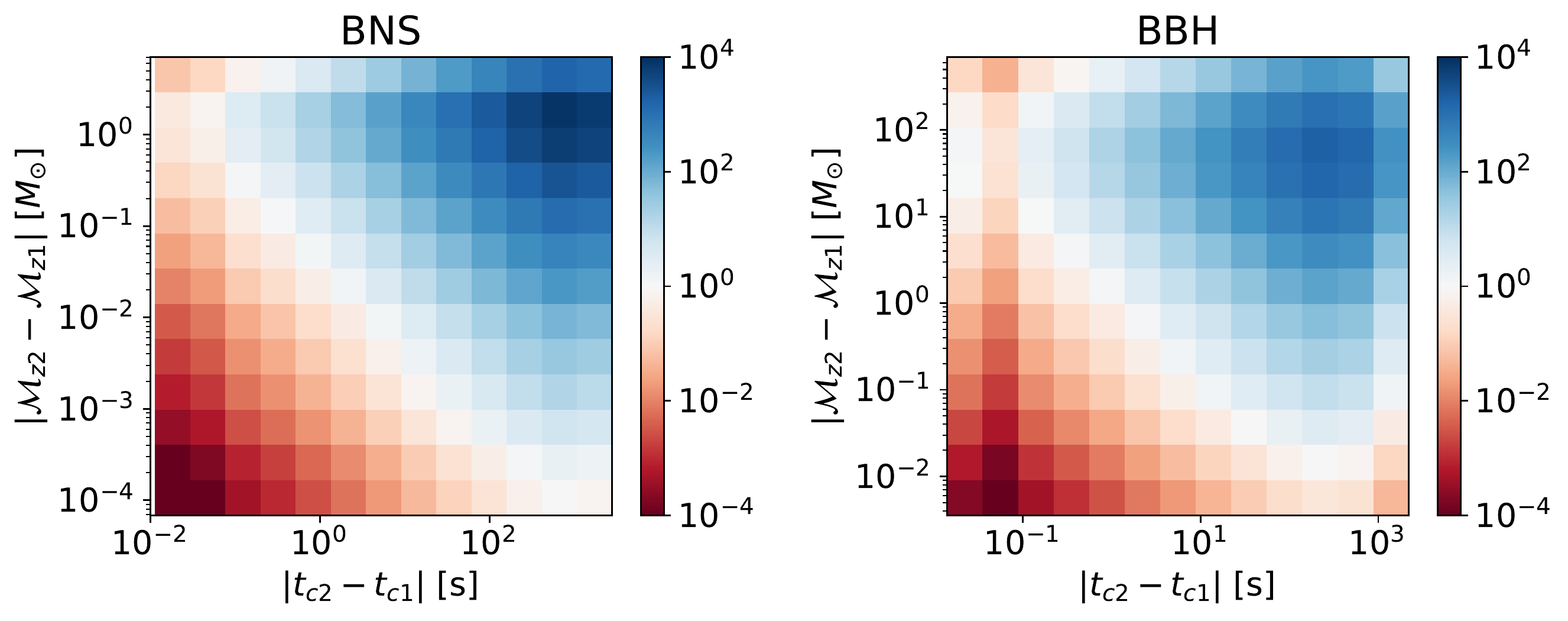}
\caption{Expected number of events per year on the plane of coalescence-time difference $|t_{\rm c2}-t_{\rm c1}|$ and redshifted chirp-mass difference $|{\cal M}_{z2}-{\cal M}_{z1}|$ for (left panel) BNSs  and (right panel) BBHs generated by assuming an event rate with a SNR$>4$. In the upper-right (blue) and lower-left (red) regions, more and less than one event per year, respectively, is expected.}
\label{fig:tc-mc-diff-dist}
\end{center}
\end{figure*}

From Fig.~\ref{fig:merger-rate-snr}, GW signals coming from BNSs and BBHs have SNRs$>1$ and $>5$, respectively. If we focus on marginally detectable events (SNR$>4$), their cumulative event rates are $\sim 2\times 10^{-3}\,{\rm s}^{-1}$ for BNSs and $\sim 6\times 10^{-4}\,{\rm s}^{-1}$ for BBHs, the latter of which generally depends on the BH masses, but the  dependence turns out to be small. With these rates, the observed number of sources is estimated by multiplying by the observation time, and the resulting values are listed in Table~\ref{tab:number-of-sources}, together with the typical duration of GW signals which are computed with the formula of the time to merger at Newtonian order \cite{Cutler:1994ys}:
\begin{equation}
t_{\rm merge}=\frac{5}{256} {\cal M}_{z} (\pi {\cal M}_{z} f )^{-8/3} \,, 
\label{eq:duration_time}
\end{equation}
where ${\cal M}_{z}=(1+z) {\cal M}$ is the redshifted chirp mass in which ${\cal M}$ is the proper chirp mass defined in the source rest frame. Note that $f$ in Eq.~(\ref{eq:duration_time}) is the lower cut-off frequency in the calculation of coalescence time.

\begin{table}[h]
\begin{center}
\begin{tabular}{cccc}
\hline \hline
Source & Per day & Per year & Signal duration \\
\hline \hline  
BNS & 172 & $6.31 \times 10^4$ & $\sim 1$ day \\
BBH & 51 & $1.89 \times 10^4$ & $\sim 1$ h \\
\hline 
\end{tabular}
\end{center}
\caption{Number of sources with a SNR$>4$.}
\label{tab:number-of-sources}
\end{table}

Based on the number of events in Table \ref{tab:number-of-sources}, we next estimate the distribution of coalescence-time difference between two adjacent signals, $|t_{\rm c2}-t_{\rm c1}|$, from the randomly generated GW events. Here, we assume that the GW events happen uniformly over time.  In Fig.~\ref{fig:tc-diff-dist}, the distributions of coalescence-time differences are plotted for one-day (left panel) and one-year (right panel) observations. In a day, the smallest time difference is typically ${\cal O}(1)\,{\rm s}$, but in a year, it becomes $\sim 10\,{\rm ms}$, a period during which the two BNS events mostly overlap, except for the last moment before merger. On the other hand, the event rate of BBHs is smaller than that of BNSs by a factor of $\approx 3$, and the resulting distribution is roughly scaled by this factor relative to the distribution of BNSs.



We also estimate the distribution of chirp-mass differences of signals close in time. Since the chirp mass is not correlated with the coalescence time, we randomly generate each mass of a binary, then compute its chirp mass according to the mass distributions of BNSs and BBHs mentioned above. Further, using the constant merger rate, we generate the redshift of a binary and compute the redshifted chirp mass that is the chirp mass actually observed in the detector frame. To be consistent with the event rate estimated for $1.33\,M_{\odot}$-$1.33\,M_{\odot}$ BNS with a SNR$>4$, we limit the redshift range of BNSs below $z=1.2$\footnote{This is not exactly correct for BNSs heavier than $1.33\,M_{\odot}$, but it does not significantly affect the results of this section.}. The resulting redshift and mass distributions normalized by the total number of events at $z<1.2$ for BNSs (upper panels) and $z<10$ for BBHs (lower panels) are shown in Fig.~\ref{fig:z-mc-dist}.
In Fig.~\ref{fig:mc-diff-dist}, for BNSs, the smallest value of the chirp-mass difference, $|{\cal M}_{z2}-{\cal M}_{z1}|$, is found to be $\sim 10^{-2}\,M_{\odot}$ in a day and $\sim 10^{-4}\,M_{\odot}$ in a year. For BBHs, it is $\sim 1\,M_{\odot}$ in a day and can become $\sim 10^{-3}\,M_{\odot}$ in a year. Relative to the BNS case, the differences  are rather large in the BBH case.  This is ascribed to the smaller event rate and wider mass distribution of BBHs than of BNSs. 

In Fig.~\ref{fig:tc-mc-diff-dist}, combining the results of redshifted chirp-mass and coalescence-time differences, we plot the expected number of events per year as a function of $|{\cal M}_{z2}-{\cal M}_{z1}|$ and $|t_{\rm c2}-t_{\rm c1}|$ for BNSs (left panel) and BBHs (right panel). Note that the results shown here are events with a SNR$>4$ for BNSs and BBHs. Figure~\ref{fig:tc-mc-diff-dist} indicates that an event with modestly small differences of coalescence times and redshifted chirp masses, roughly $|{\cal M}_{z2}-{\cal M}_{z1}||t_{\rm c2}-t_{\rm c1}|\sim 10^{-1}$ $M_\odot\,\,$s for BNSs and  $\sim 1$ $M_\odot\,\,$s for BBHs, can happen in a year of  observation. In the next section, based on this result, we investigate the impacts of closely overlapping signals on the parameter estimation. 


\section{Fisher matrix formalism}
\label{sec:Fisher_Matrix}
In this section, based on the Fisher matrix formalism, we present the prescription to estimate the statistical errors of the parameters in the overlapping GW signals (Sec.~\ref{subsec:Formalism}). Further, the systematic bias in the parameter estimation, caused by a misinterpretation of the overlapping signal, is described. In Sec.~\ref{subsec:analytical}, the condition that the impact of the overlapping signal becomes significant is analytically investigated, and the criteria to potentially produce a larger statistical error are derived at the Newtonian order. 

Throughout the analysis, we consider the inspiral GW waveform, dropping the merger and ringdown parts. This is because SNR and the statistical errors of parameters for relatively low-mass BBH with redshifted chirp mass $\lesssim 40\,M_{\odot}$ are predominantly determined by the contribution from the low-frequency band, and neglecting the merger-ringdown part of the waveform does not give a large impact on the parameter estimation.
For relatively massive BBH, the contribution from the merger-ringdown part of the waveform is crucial in the Fisher analysis, since it dominates the one from the inspiral part of the waveform. However, Fig.~\ref{fig:z-mc-dist}  shows that BBH with redshifted chirp mass beyond $50 M_{\odot}$ are rather rare, and the occurence rate of such closely overlapping events becomes even small. That is, those events would not be a serious concern, and hence focusing on the inspiral phase, we  estimate the statistical errors for relatively low-mass binary systems (see Table \ref{tab:binary_parameters}).

 For the inspiral waveform, we take the post-Newtonian (PN) terms up to the second order and present mainly the results for the overlapping signals at the Newtonian order and second PN (2PN) order. The main reason for this is that the number of parameters no longer increases beyond the 2PN order, and including the higher-order PN corrections does not significantly change the results. 

The Fourier transform of the signal coming from a single binary system, $h(t)$, is given by ~\cite{Cutler:1994ys,Maggiore:book,Poisson:1995ef}
\begin{align}
\tilde{h}(f) &=  \mathcal{A} \, f^{-7/6} e^{i \Phi(f)} \;, \\
\mathcal{A}  &=  \frac{1}{\sqrt{30} \pi^{2/3}}\frac{ {\cal M}_{z }^{5/6} }{d_{\rm L}(z)}\;, 
\label{eq:single-event_signal}
\end{align}
where $d_{\rm L}$ is the luminosity distance.
Here, we consider the sky-averaged waveform with aligned spins. The phase function $\Phi$ calculated is expressed as 
\begin{align}
\Phi (f) &= 2\pi f \,t_{\rm c} -\phi_{\rm c} -\frac{\pi}{4} +\frac{3}{128} 
(\pi {\cal{M}}_{z } f)^{-5/3} \nonumber \\
& \times \left[ 1 + \frac{20}{9} \left( \frac{743}{336}+\frac{11}{4} \eta \right) x -4(4\pi-\beta) x^{3/2} \right. \nonumber \\
& \left. + \left( \frac{15293365}{508032}+ \frac{27145}{504} \eta + \frac{3085}{72} \eta^2 -10 \sigma \right) x^2 \right] \;, 
\label{eq:gw-phase}
\end{align}
where $t_{\rm c}$ and $\phi_{\rm c}$ are the time and phase at coalescence, respectively, $\eta$ is the symmetric mass ratio,  $\beta$ is the spin-orbit coupling parameter, $\sigma$ is the spin-spin coupling parameter, and $x \equiv (\pi M_{z} f )^{2/3}$ with the redshifted total mass given by $M_{z}={\cal{M}}_{z } \eta^{-3/5}$. 
In Eq.~(\ref{eq:gw-phase}), the square brackets describe the post-Newtonian corrections expanded in powers of $x$; that is, according to the power of  $x$, each term represents the Newtonian, 1PN, 1.5PN, 2PN, in order.  To sum up, a single GW signal at the 2PN order has seven parameters:
\begin{align}
\bm{\theta} = (\ln \mathcal{A}, t_{\rm c}, \phi_{\rm c}, \ln {\cal{M}}_{z},  \ln \eta, \beta, \sigma). 
\label{eq:gw-parameters}
\end{align}
Note that the number of parameters doubles in the case of an overlapping signal, i.e., from 4 to 8 at the Newtonian, and from 7 to 14 at the 2PN order.

\subsection{Estimating statistical errors and systematic biases in overlapping signals}
\label{subsec:Formalism}

\subsubsection{Statistical errors in overlapping signals}
\label{subsubsec:statistical errors in overlapping signals}

First consider first the parameter estimation in a single GW event. Given the waveform of the template $\tilde{h}(f)$ and noise spectral density $S_{\rm{n}}$, the Fisher matrix is given by (see, e.g., \cite{Cutler:1994ys,Poisson:1995ef})
\begin{equation}
\Gamma_{ab} = 4 \, {\rm{Re}}  \int_{f_{\rm{min}}}^{f_{\rm{max}}}
 \frac{\partial_{a} \tilde{h}^{\ast}(f)\, \partial_{b}
 \tilde{h}(f)}{S_{\rm{n}}(f)} df \;,
\label{eq:Fisher-matrix} 
\end{equation}
where the symbol $\partial_a$ stands for the derivative with respect to a parameter $\theta_a$ and $h^{\ast}$ is the complex conjugate of $h$. The range of integral $[f_{\rm min}, f_{\rm max}]$ is restricted to the observed frequency range. The frequency $f_{\rm{max}}$ is chosen to be twice the frequency of the innermost stable circular orbit, $f_{\rm ISCO}=(6^{2/3}\pi M_{z})^{-1}$. 
As the minimum frequency $f_{\rm{min}}$ is determined by the lower cutoff of the noise spectral density, we adopt $f_{\rm{min}}=1$ Hz.

Provided the Fisher matrix, the statistical error of a parameter marginalized over others, which we denote by $\delta\theta_a$, is estimated to be 
\begin{equation}
\delta \theta_{a}=\sqrt{(\Gamma^{-1})_{aa}},
\label{eq:Inverse-Fisher-matrix} 
\end{equation}
where the matrix $(\Gamma^{-1})_{ab}$ is the inverse of Fisher matrix. 

Next we consider the statistical errors for overlapping GW signals. In this case, 
the Fourier transform of the GW signal, $\tilde{h}(f)$, is made of the superposition of the two waveforms given by
\begin{align}
\tilde{h}(f) &=  \sum_{j=1}^{2} \mathcal{A}_{j} \, f^{-7/6} e^{i \Phi_{j}(f)} \;, 
\label{eq:overlap_signal} \\
\mathcal{A}_{j}  &=  \frac{1}{\sqrt{30} \pi^{2/3}}\frac{ {\cal M}_{z j}^{5/6} }{d_{\rm L}(z_j)}\;,
\end{align}
where $\Phi_{j}$ is the phase function in Eq.~(\ref{eq:gw-phase}) with a set of parameters $\{\bm{\theta}_j\}$. The Fisher matrix for the overlapping signals is simply obtained by substituting Eq.~(\ref{eq:overlap_signal}) into Eq.~(\ref{eq:Fisher-matrix}), with the number of parameters doubled (and hence the number of matrix elements squared). Thus, the structure of the Fisher matrix is mostly similar to the one in the single-event case, but a notable difference is the off-diagonal blocks involving the term, $\partial_{a} \tilde{h}_{1}^{\ast}\, \partial_{b} \tilde{h}_{2}$, which characterizes the interference between the two signals. The presence of this interference can induce the parameter degeneracy between the GW signals 1 and 2, and the statistical errors potentially get increased. 
In Sec.~\ref{subsec:analytical}, we analytically estimate at the Newtonian order the impact of this interference, and derive the conditions at which the interference becomes significant.

\subsubsection{Biased parameter estimation due to the misinterpretation of overlapping signals}
\label{subsubsec:systematic_bias}

Given the likelihood function, the Fisher matrix formalism also provides a simple way
to estimate the biases in the best-fit parameters caused by an incorrect template. For the overlapping events for our interest, what is likely to occur is that the signal-to-noise ratio for one of the overlapping events will be small, and we misinterpret the fact that the detected signal is made of a single GW event, ignoring the event with a small signal-to-noise ratio in the parameter estimation analysis. This can potentially affect the best-fit parameter for one of the GW events considered.

Consider an overlapping event signal made of two merger events, 1 and 2. If one misinterprets the overlapping signal as a single GW event, and tries to estimate only the parameters of event 1, the best-fit parameter for 
event 1 can systematically deviate from the true value $\theta_{1 a}^{\rm true}$ due to the wrong assumption of the GW template ignoring event 2. We thus obtain the biased best-fit parameters $\theta_{1 a}^{\rm true}\to \theta_{1 a}^{\rm true}+ \Delta\theta_{1 a}$, and the bias $\Delta\theta_{1 a}$ is estimated from 
\begin{align}
\Delta \theta_{1 a}&=\sum_{b} \bigl(\mathcal{F}^{-1}\bigr)_{ab}\,s_b    
\label{eq:formula_systematic_bias}
\end{align}
with the matrix $\mathcal{F}_{ab}$ and the vector $s_a$, respectively, given by
\begin{align}
\mathcal{F}_{ab} &= 4\,{\rm Re}\int_{f_{\rm min}}^{f_{\rm max}} \frac{\partial_a \tilde{h}^*_1(f)\partial_b \tilde{h}_1(f)-\tilde{h}_2^*(f)\,\partial_a\partial_b \tilde{h}_1(f)}{S_{\rm{n}}(f)}\,df,
\label{eq:matrix_F}
\\
s_a&=4\,{\rm Re}\int_{f_{\rm min}}^{f_{\rm max}}  \frac{\tilde{h}_2^*(f)\partial_a\tilde{h}_1(f) }{S_{\rm{n}}(f)}\,df.
\label{eq:vector_s}
\end{align}
Derivation of this expression is presented in the Appendix \ref{sec:derivation}. Note that the integrands in these expressions are evaluated at the fiducial (true) parameters. Equation~(\ref{eq:formula_systematic_bias}) includes the contributions coming from event 2, which is ignored in the parameter estimation analysis. For a low signal-to-noise ratio of event 2, we have $|\tilde{h}_2|\ll|\tilde{h}_1|$ and, in such a case, the matrix $\mathcal{F}_{ab}$ is reduced to the Fisher matrix for the single event 1, $\Gamma_{ab}$. Then the size of the systematic bias $\Delta\theta_a$ is linearly proportional to the ratio of the GW amplitudes, and hence the ratio of the SNR defined at Eq.~(\ref{eq:SNR}).

\subsection{Analytical estimation}
\label{subsec:analytical}


Based on the formalism in Sec.~\ref{subsec:Formalism}, 
we analytically estimate the condition under which the impact of the overlapping signals can be ignored, focusing specifically on their statistical errors. As we previously mentioned, in the Fisher matrix at Eq.~(\ref{eq:Fisher-matrix}) with $\tilde{h}$ given by Eq.~(\ref{eq:overlap_signal}),  the off-diagonal blocks involving the term $\partial_a\tilde{h}_1^*\partial_b\tilde{h}_2$ are responsible for the interference between two signals,
\begin{equation}
\int\frac{\partial_{a} \tilde{h}_1^{\ast}\, \partial_{b}
 \tilde{h}_2}{S_{\rm{n}}}\, df= \mathcal{A}_1 \mathcal{A}_2  \int f^{-7/3} \frac{\partial_{a} \Phi_1^{\ast}\, \partial_{b}
 \Phi_2}{S_{\rm{n}}} e^{i(\Phi_2-\Phi_1)}\, df\;.
 \label{eq:interference_Fisher}
 \end{equation}
If these blocks become large, the parameters in the GW events 1 and 2 are not independently estimated, and there appears to be a certain amount of correlation, potentially leading to an increased error. Thus, for the impact of overlapping signals to be negligible, these off-diagonal blocks must be sufficiently small relative to the diagonal blocks, in particular, the diagonal components of the Fisher matrix. This gives  
\begin{align}
& \mathcal{A}_2 \left| \int d \ln f f^{-4/3} \frac{\partial_{a} \Phi_1^{\ast}\, \partial_{b}
 \Phi_2}{S_{\rm n}(f)} \cos (\Phi_2-\Phi_1) \right| \nonumber \\
& \qquad \qquad \qquad \qquad \ll \mathcal{A}_1\int d \ln f f^{-4/3} \frac{|\partial_{a} \Phi_1|^2}{S_{\rm n}(f)} \;.
\label{eq:inequality_Fisher_matrix}
\end{align} 
An explicit evaluation of the integrals on both sides of Eq.~(\ref{eq:inequality_Fisher_matrix}) needs functional forms of $\Phi_j$ and $S_{\rm n}$. However, the integrand on the left-hand side involves an oscillating function with a negative power of frequency $f^{-4/3}$. Thus, as long as the 
noise spectral density is smooth enough relative to the waveform, the left-hand side is dictated mostly by the integral near the low-frequency cutoff $f\sim f_{\rm min}$. In this respect, one would expect the inequality in Eq.~(\ref{eq:inequality_Fisher_matrix}) to hold if the phase difference $\Phi_2-\Phi_1$ is sufficiently large at the lower frequency cutoff. That is, 
\begin{align}
\bigl|\Phi_2-\Phi_1\bigr|_{f=f_{\rm min}} \gg \pi \;.
\label{eq:non-overlapping-condition0}
\end{align}    

The condition in Eq.~(\ref{eq:non-overlapping-condition0}) is fairly general and we do not need an explicit functional form of $\Phi$. For more explicit calculation, we consider the Newtonian waveform below. We have
 \begin{align}
 \Phi_2-\Phi_1 &= 2 \pi f (t_{\rm c2}-t_{\rm c1}) - (\phi_{\rm c2}-\phi_{\rm c1}) \nonumber \\
 &+\frac{3}{128} (\pi f)^{-5/3}  \left({\cal M}_{z2}^{-5/3} - {\cal M}_{z1}^{-5/3}  \right) \;.
\end{align}
Ignoring the frequency independent phase factor, Eq.~(\ref{eq:non-overlapping-condition0}) is reduced to
\begin{align}
&\biggl| 2 f_{\rm min} (t_{\rm c2}-t_{\rm c1})  \nonumber \\
& + \frac{3}{128} \pi^{-8/3} f_{\rm min}^{-5/3} \left( {\cal M}_{z2}^{-5/3} - {\cal M}_{z1}^{-5/3} \right) \biggr| \gg 1\;. 
\label{eq:non-overlapping-condition}
\end{align}

\begin{table*}[htb!]
\begin{ruledtabular}
\caption{Parameters of binary systems used in the Fisher matrix analysis
\label{tab:binary_parameters}}
\begin{tabular}[c]{c|cccccc} 
Binary type & $m_{1}$ & $m_{2}$ & $\mathcal{M}$ & SNR(1ET) & $z$ & $\mathcal{M}_{z}$ 
   \\ & ($M_{\odot}$) & ($M_{\odot}$)  & ($M_{\odot}$) & \,\, & &($M_{\odot}$)  \\
   \hline 
   BNS 
   & $1.33$  & $1.33$ & $1.15$ & $5$ & $0.92$ & $2.2$ \\
   BBH1 
   & $10$  & $10$ & $8.71$ & $15$ & $1.67$ & $23.3$ \\
   BBH2 
   & $20$  & $20$ & $17.4$ & $80.3$ & $0.44$ & $25.0$ \\
   BBH3 
   & $10$  & $10$ & $8.71$ & $13.8$ & $1.87$ & $25.0$ \\
  \end{tabular}
\end{ruledtabular}
\end{table*}

\begin{table*}[htb!]
\begin{ruledtabular}
\caption{Expected errors on the parameters of binary systems in the case of single-event GWs. The results are derived based on the Fisher matrix analysis assuming a single ET-like detector and spinless system (i.e., $\beta=0=\sigma$). Note that the expected errors on the parameters of BBH3 in Table \ref{tab:binary_parameters} are obtained from those of the BBH2 when multiplying by the factor of $(80.3/13.8)\simeq5.82$.
\label{tab:single_event_errors}}
\begin{tabular}[c]{c|rrrrrrr} 
Binary type   & PN order & $ \delta t_{{\rm c}}$ \,\,& $\delta \phi_{{\rm c}}$ \,\,& $\delta \mathcal{M}_{z}/\mathcal{M}_{z}$ & $\delta \eta/\eta $ & $\delta \beta$ & $\delta \sigma$   
   \\ & & (ms)  & (rad) & (\%) \,\,\,\, & (\%)\,\,\,\, \\ \hline 
   BNS 
   & Newtonian  & 0.393 & 0.330 & 0.00100 & $\cdots$ & $\cdots$ & $\cdots$ 
     \\
  &  1 &0.484&0.616&0.00370&1.23& $\cdots$ & $\cdots$ 
    \\ 
& 1.5 &0.740&2.09&0.0111&10.4& 0.503 & $\cdots$ 
    \\ 
  &    2  & 1.65 & 18.2 &0.0290 & 49.9 & 1.21 & 6.69 
   \\ \hline 
  BBH1 
  &  Newtonian  & $0.703$ & $0.173$ & $0.0193$ & $\cdots$ & $\cdots$ & $\cdots$ 
     \\
 &   1 & 1.28 & 0.534 & 0.0858 & 7.67 & $\cdots$ & $\cdots$ 
    \\ 
& 1.5 & 3.36 & 3.10 & 0.339 & 87.3 & 5.31 & $\cdots$  
    \\ 
  &    2  & 13.8 & 48.3 & 1.33 & 675 & 31.3 & 21.0   
     \\ \hline 
  BBH2 
  &  Newtonian  & 0.142 & 0.0333 & 0.00409 & $\cdots$ & $\cdots$ & $\cdots$
     \\
 &   1 & 0.264 & 0.105 & 0.0183 & 1.59 & $\cdots$ & $\cdots$ 
    \\ 
 &   1.5 & 0.706 & 0.624 & 0.0736 & 18.4 & 1.12 & $\cdots$ 
    \\ 
 &   2 & 2.96 & 9.94 & 0.295 & 144 & 6.76 & 4.30 
    \\ 

  \end{tabular}
\end{ruledtabular}
\end{table*}

In Eq.~(\ref{eq:non-overlapping-condition}), there are two limiting cases in which the impact of the overlapping signal is negligible. One is the case in which the coalescence-time difference is large. Equation~(\ref{eq:non-overlapping-condition}) is then simplified to give
\begin{equation}
|t_{\rm c2}-t_{\rm c1}|  \gg \frac{1}{2 f_{\rm min}} \approx 5.0 \times 10^{-2} \left( \frac{10\,{\rm Hz}}{f_{\rm min}} \right)\,\rm [s] \;. 
\label{eq:non-overlapping-condition1}  
\end{equation} 
Here, the minimum frequency $f_{\rm min}$ is taken to be $f_{\rm min}=10\,{\rm Hz}$. This roughly corresponds to the cutoff frequency below which the noise spectral density rapidly goes up due to the seismic noise.  Another limiting case appears to manifest when the coalescence-time difference is small. In this case, the second term on the left-hand side of Eq.~(\ref{eq:non-overlapping-condition}) becomes dominant, and the condition is  reduced to  
\begin{align}
|\Delta Q| & \gg \frac{128 \pi^{8/3}}{5} (f_{\rm min} {\cal M}_{z1})^{5/3} \nonumber \\
& \approx 1.3 \times 10^{-4} \left( \frac{{\cal M}_{z1}}{2.2\,M_{\odot}} \right)^{5/3} \left( \frac{f_{\rm min}}{10\,{\rm Hz}} \right)^{5/3} \;,
\label{eq:non-overlapping-condition2}
\end{align}
where the quantity $\Delta Q$ is the fractional chirp-mass difference defined by 
$\Delta Q \equiv ({\cal M}_{z2} - {\cal M}_{z1})/{\cal M}_{z2}$.
In deriving Eq.~(\ref{eq:non-overlapping-condition2}), we assumed that $\Delta Q$ is small ($|\Delta Q|\ll1$), which can be realized in most cases when the impact of the overlapping signals becomes non-negligible. Thus, the condition (\ref{eq:non-overlapping-condition2}) implies that $|\Delta Q|$ must not be too small.

\begin{figure*}[htb!]
\begin{center}
\includegraphics[width=15cm,angle=0,clip]{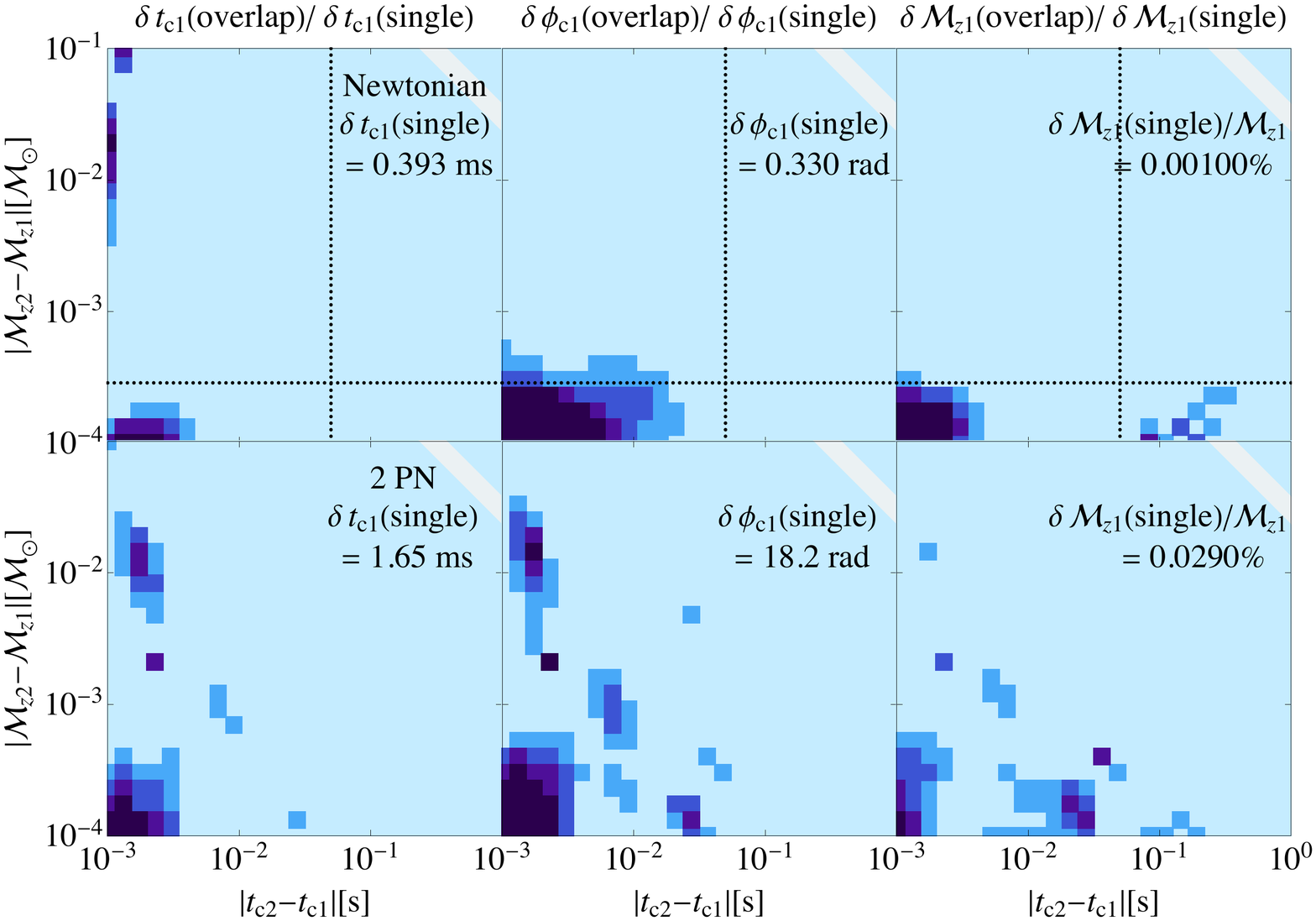}
\hspace{-5mm}
\mbox{\raisebox{1.5cm}{\includegraphics[height=7.5cm, angle=0,clip]{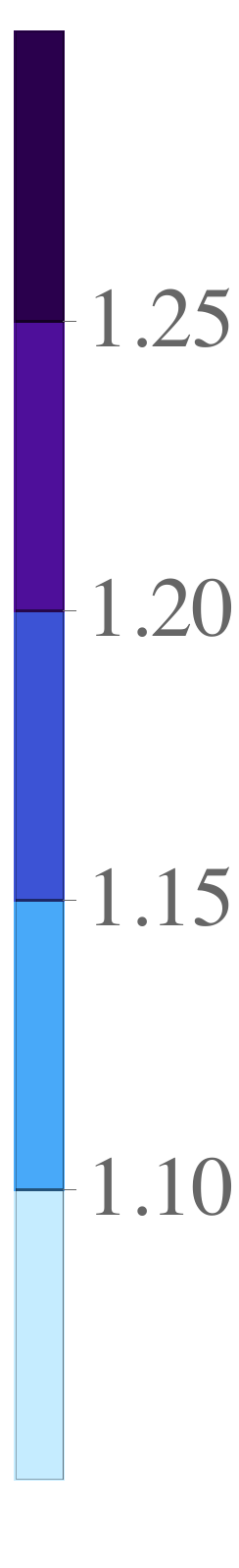}}}
\end{center}
\vspace*{-.5cm}
\caption{Expected errors on (left panels) the coalescence time,  (middle panels) phase, and  (right panels) redshifted chirp mass of the BNS system in overlapping GW events, under the assumption that each of the overlapping signals is spinless (i.e., $\beta=0=\sigma$) with synchronized phases, $\phi_{\rm c1}=\phi_{\rm c2}=0$. The resulting errors estimated at the (upper panels) Newtonian and (lower panels) 2PN orders are normalized by those estimated in the single-event BNS case listed in Table \ref{tab:single_event_errors}. Here, the two GW signals composing the BNS and BNS$^\prime$ systems, whose parameters are almost identical except for the coalescence time and redshifted chirp masses (see Table \ref{tab:binary_parameters}), are assumed to overlap over time. Then the statistical errors on the parameters for one of the overlapping systems are estimated and plotted as a function of the differences of the coalescence time and redshifted chirp masses, i.e., $|t_{\rm c2}-t_{\rm c1}|$ and $|\mathcal{M}_{z2}-\mathcal{M}_{z1}|$, in the $26\times26$ pixelized plane, with the results evaluated at the center of each pixel. For reference, the expected error in each parameter in the case of the single-event GWs is indicated in each panel. Furthermore, white stripes indicate the boundary below which the expected number of overlapping events is less than one per year (see Fig.~\ref{fig:tc-mc-diff-dist} ).
In the upper panels, the vertical and horizontal dotted lines, respectively, represent the critical conditions given in 
Eqs.~(\ref{eq:non-overlapping-condition1}) and  (\ref{eq:non-overlapping-condition2}), which are analytically derived in Sec.~\ref{subsec:analytical}.}
\label{fig:Contour_Error_BNS}
\end{figure*}

The inequality in Eq.~(\ref{eq:non-overlapping-condition0}) or, equivalently, 
either Eq.~(\ref{eq:non-overlapping-condition1}) or Eq.~(\ref{eq:non-overlapping-condition2}) gives a sufficient condition to mitigate the impact of overlapping signals in most cases. A subtlety arises when we take the components $a$ and $b$ to be $t_{\rm c1}$ and $t_{\rm c2}$ in Eq.~(\ref{eq:inequality_Fisher_matrix}).  In this case, the integrand on the left-hand side of Eq.~(\ref{eq:inequality_Fisher_matrix}) has a positive power in frequency and, in contrast to the cases that we considered above, the integral is dominated by the behaviors of the integrand at $f\sim f_{\rm max}$. Thus, one may need an additional condition to mitigate the impact of the overlapping signals, i.e.,  $|\Phi_2-\Phi_1|_{f=f_{\rm max}} \gg \pi $. In cases with the Newtonian waveform, this gives  Eq.~(\ref{eq:non-overlapping-condition}), with $f_{\rm min}$ replaced by $f_{\rm max}$. As we discussed above, we can similarly consider the two limiting cases. However, since $f_{\rm max}\gg f_{\rm min}$, only the following condition is to be considered: 
\begin{equation}
|t_{\rm c2}-t_{\rm c1}|  \gg \frac{1}{2 f_{\rm max}} \approx 5.0 \times 10^{-4} \left( \frac{1\,{\rm kHz}}{f_{\rm max}} \right)\,\rm [s] \;. 
\label{eq:non-overlapping-condition3}
\end{equation}
This is simply the time resolution of a signal processing and is actually weaker than Eq.~(\ref{eq:non-overlapping-condition1}).

In summary, if either of Eq.~(\ref{eq:non-overlapping-condition1}) or Eq.~(\ref{eq:non-overlapping-condition2}) is fulfilled, the off-diagonal blocks between signals in the Fisher matrix can be ignored and the overlap between signals does not affect the parameter estimation. Although we focus here on the Fisher matrix for the statistical error estimation, the interference term between two signals like Eq.~(\ref{eq:interference_Fisher}) also appears in the expressions for the systematic bias estimation [see Eqs.~(\ref{eq:matrix_F}) and (\ref{eq:vector_s})].
Thus, one expects that the same conditions as derived above would hold in order to mitigate the systematic biases. 
Furthermore, as long as the PN corrections are small, the conditions still remain relevant to the PN waveform.  Finally, note that Refs.~\cite{Samajdar:2021egv, Pizzati:2021gzd} investigated 
the parameter regions satisfying Eqs.~(\ref{eq:non-overlapping-condition1}) or (\ref{eq:non-overlapping-condition2}), and they consistently obtained results indicating that the overlapping signals do not have a serious impact on the parameter estimation. In the next section, we will also check this against the Fisher matrix analysis, including the regime where the conditions at  Eqs.~(\ref{eq:non-overlapping-condition1}) and  (\ref{eq:non-overlapping-condition2}) are not satisfied.

\section{Results of the Fisher Forecast}
\label{sec:results}

\begin{figure*}[tb]
\begin{center}
\includegraphics[width=15cm,angle=0,clip]{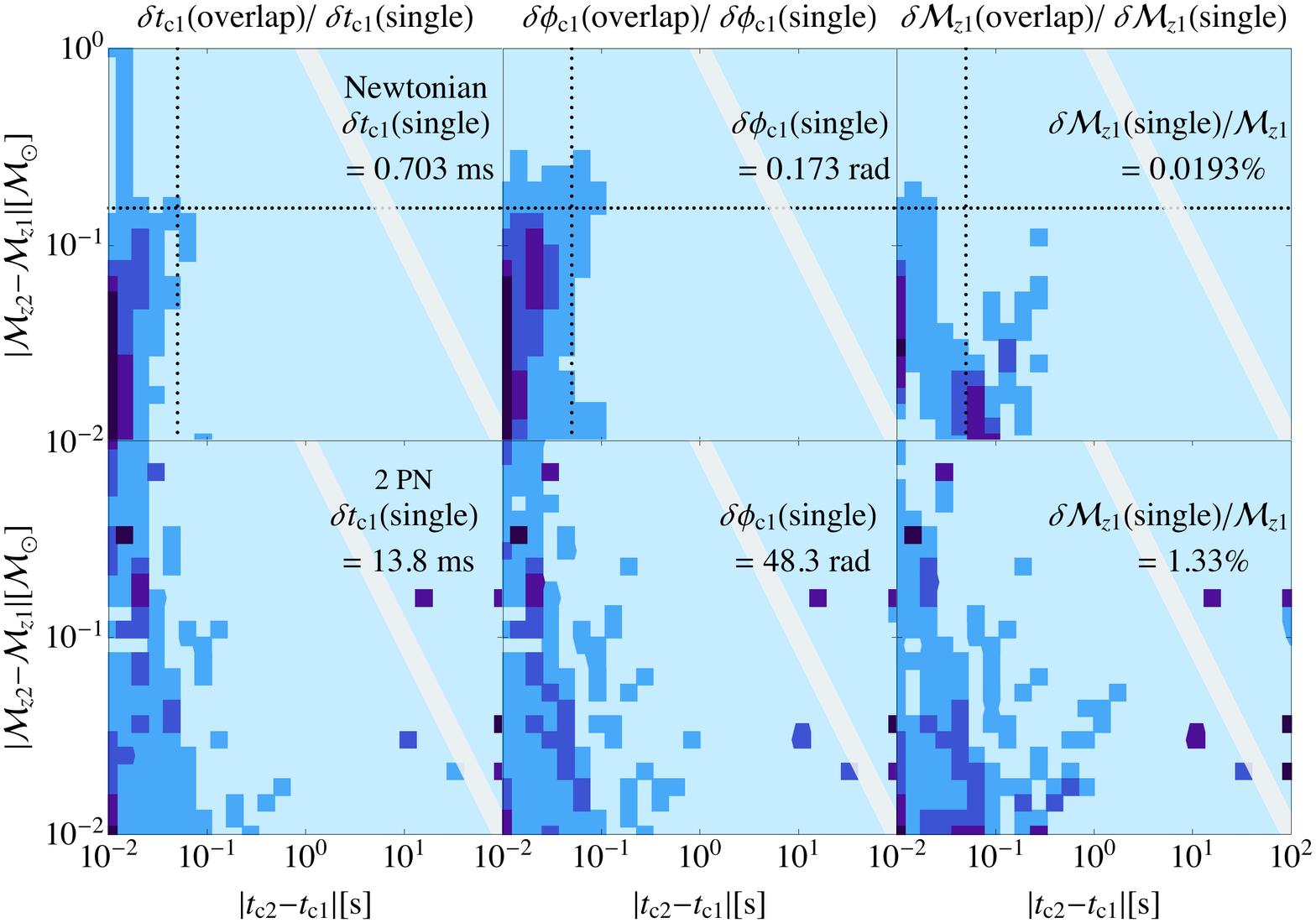}
\hspace{-5mm}
\mbox{\raisebox{1.5cm}{\includegraphics[height=7.5cm, angle=0,clip]{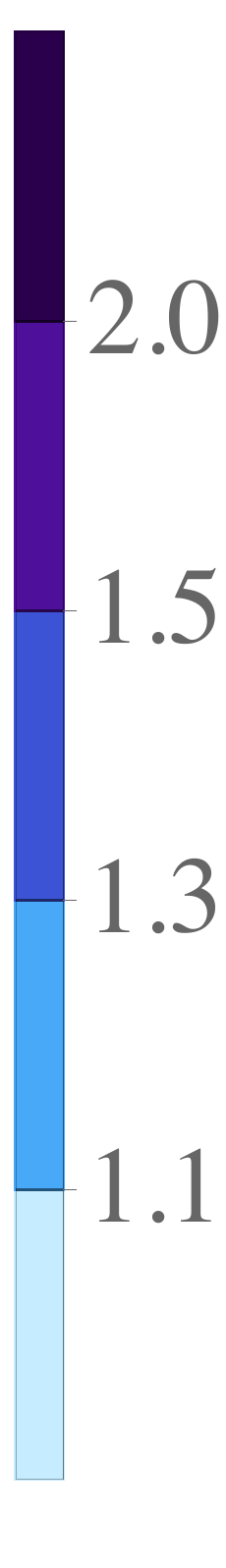}}}
\end{center}
\vspace*{-.5cm}
\caption{Same as Fig.\ref{fig:Contour_Error_BNS}, but for the overlapping BBH events, assuming the parameters of the BBH1 system listed in Table \ref{tab:binary_parameters}. Note that the color scale adopted here is different than the one in  Fig.~\ref{fig:Contour_Error_BNS}.}

\label{fig:Contour_Error_BBH}
\end{figure*}

\begin{figure*}[t]
\begin{center}

\includegraphics[width=7cm,angle=0,clip]{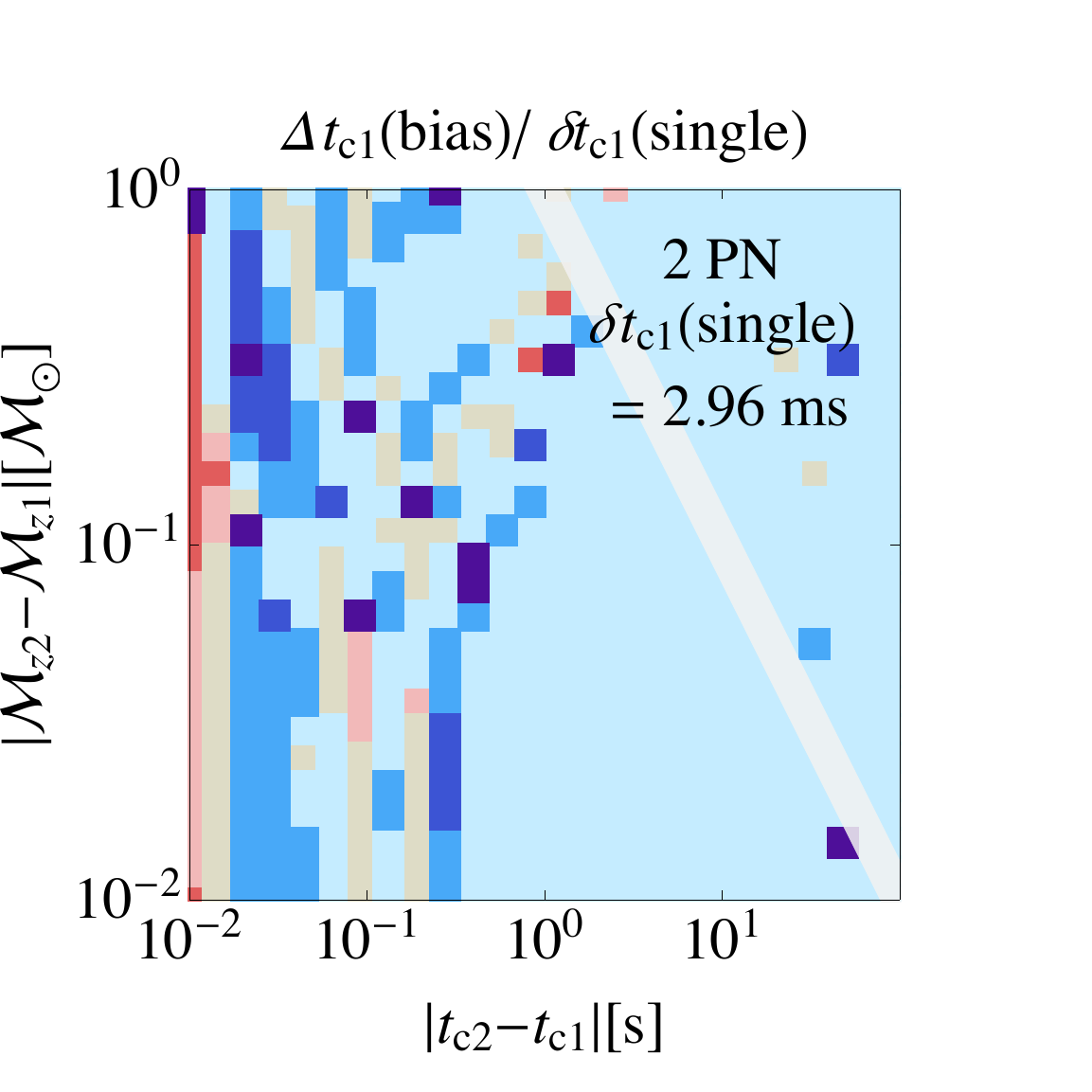}
\hspace{-2.66cm}
\includegraphics[width=7cm,angle=0,clip]{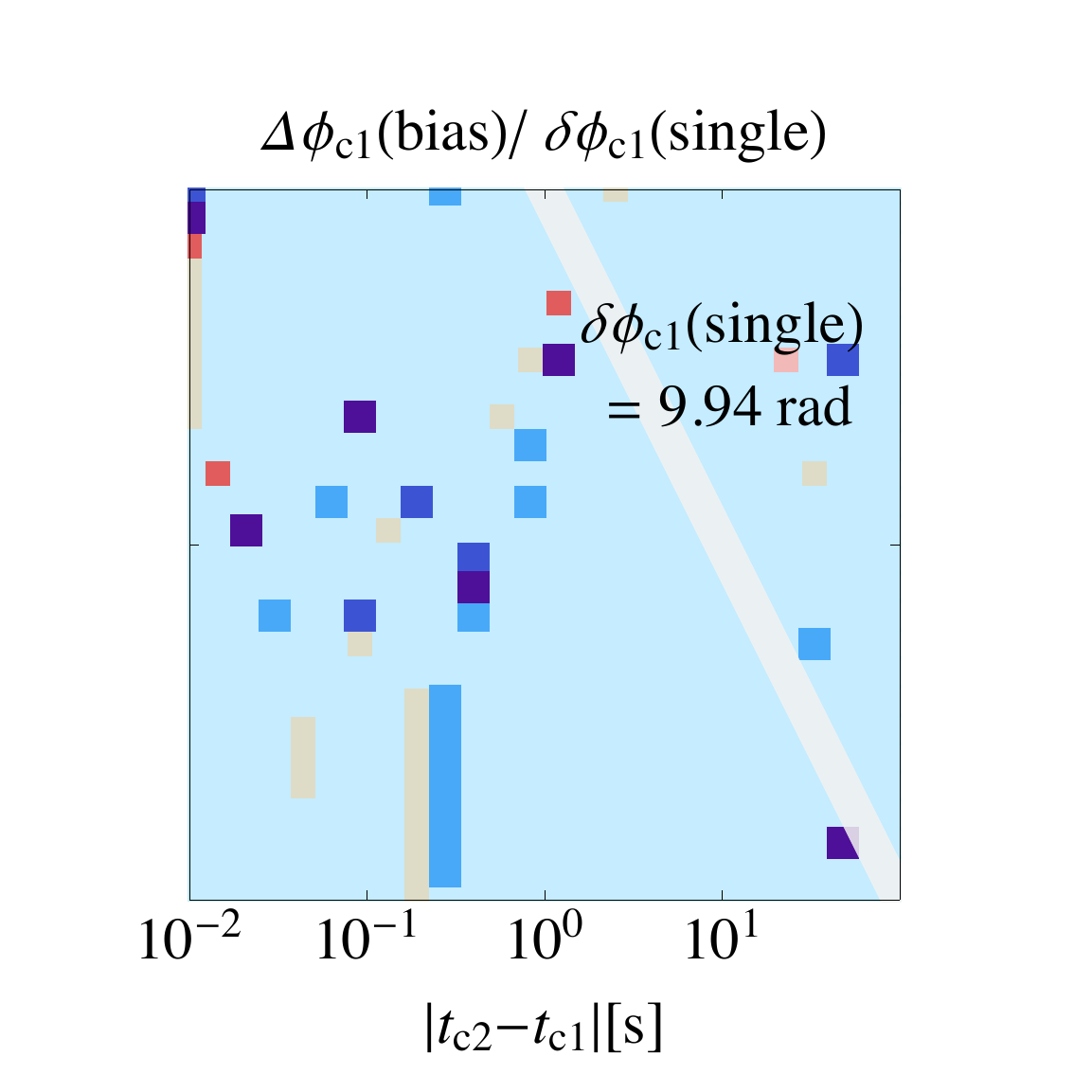}
\hspace{-2.66cm}
\includegraphics[width=7cm,angle=0,clip]{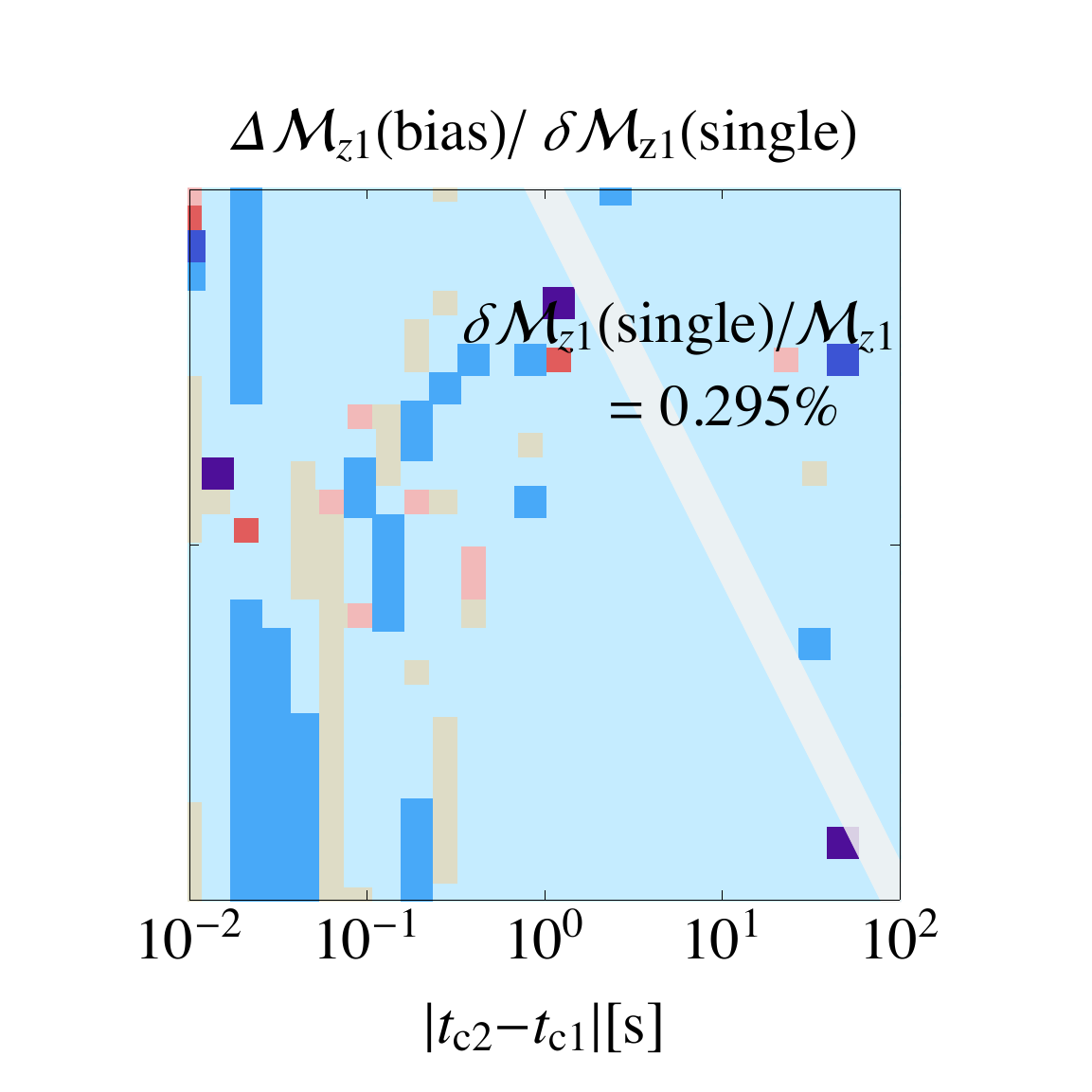}
\hspace{-1cm}
\mbox{\raisebox{0.95cm}{\includegraphics[height=4.95cm, angle=0,clip]{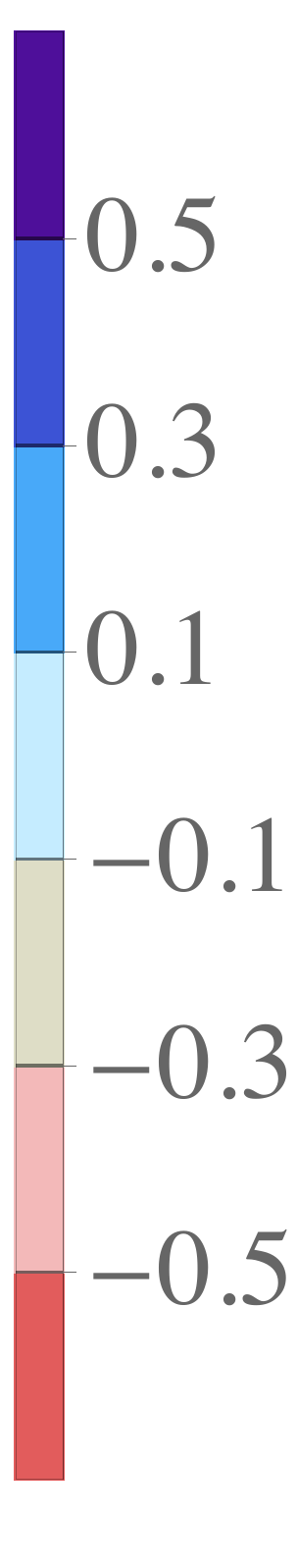}}}
\end{center}
\vspace*{-.5cm}
\caption{Systematic biases in the best-fit value of  (left panel) the coalescence time, (middle panel) the phase, and  (right panel) the redshifted chirp mass estimated from the Fisher matrix formalism in Sec.~\ref{subsubsec:systematic_bias}. Here, we consider the overlapping GW event coming from the BBH2 and BBH3 systems listed in Table \ref{tab:binary_parameters}. Erroneously ignoring the signal from BBH3, the size of the biased parameter estimation for the BBH2 system is computed and the results are then normalized by the statistical errors inferred from the single-event case. The two-dimensional $26\times26$ pixelized plot shown in each panel summarizes the results of the systematic bias for which the coalescence time and redshifted chirp mass for the BBH3 system are slightly shifted from Table \ref{tab:binary_parameters}. For reference, the statistical error inferred from the single-event case is also indicated in each panel. In each panel, the white stripe represents the boundary below which the expected number of overlapping GW signals is less than one per year.
}
\label{fig:Contour_Bias}
\end{figure*}


Based on the results of the Monte Carlo analysis in Sec.~\ref{sec:parameter-dist}, we concerned ourselves with overlapping GW events in which the parameters of each binary system, in particular, the coalescence time $t_{\rm c}$ and redshifted chirp mass $\mathcal{M}_z$, are very close to each other. As shown in Fig.~\ref{fig:tc-mc-diff-dist}, there are a certain amount of occurrences of such events expected, potentially having a large impact not only on the parameter estimation study of each event but also, indirectly, on the detection of stochastic GW backgrounds through the foreground noise subtraction.

To quantitatively investigate the impact of those events, we consider below the four representative binary systems summarized in Table \ref{tab:binary_parameters}, together with an estimated SNR based on
\begin{equation}
{\rm SNR}^2 =4  \int_{f_{\rm{min}}}^{f_{\rm{max}}} \frac{|\tilde{h}(f)|^2}{S_{\rm{n}}(f)} df \;,
\label{eq:SNR}
\end{equation}
with $f_{\rm min}=1$ Hz and $f_{\rm max}=f_{\rm ISCO}=(6^{2/3}\pi\,M_z)^{-1}$. 
We then examine the Fisher matrix analysis and compute the statistical errors on the parameters of overlapping BNS-BNS$^\prime$ and BBH1-BBH1$^\prime$ systems. Here, the binary system BNS$^\prime$ (BBH1$^\prime$) is almost identical to BNS (BBH1) except for the coalescence time $t_{\rm c}$ and redshifted chirp mass $\mathcal{M}_z$.
Also, the systematic biases in the parameter estimation are computed for the overlapping GWs consisting of BBH2 and BBH3, the latter of which is the event with a low SNR and is erroneously ignored in the parameter estimation analysis. 
In Table \ref{tab:single_event_errors}, for reference, the expected errors on the parameters of each binary system are estimated, assuming that it is a single spinless (i.e., $\beta=0=\sigma$) GW event without any overlapping. Note that throughout the Fisher matrix analysis, we consider a single ET-like detector.

The results of our Fisher matrix analysis are presented in Figs.~\ref{fig:Contour_Error_BNS}--\ref{fig:Contour_Bias}, all of which are shown as a function of the differences of the coalescence time, $|t_{\rm c2}-t_{\rm c1}|$, and the differences of the redshifted chirp mass, $|\mathcal{M}_{z2}-\mathcal{M}_{z1}|$, focusing particularly on the estimations of the coalescence time (left panels),  the phase (middle panels), and the redshifted chirp mass (right panels). For reference, we also plot the boundary below which the expected number of overlapping events, estimated in Sec.~\ref{sec:parameter-dist}, is less than one per year (see Fig.~\ref{fig:tc-mc-diff-dist}; depicted as a white stripe). Note that  
the forecast results are obtained by setting $t_{\rm c1}$ to zero and varying $t_{\rm c2}$, with phases of the GW signals at  synchronized coalescence times,  i.e., $\phi_{\rm c1}=0=\phi_{\rm c2}$.  For the redshifted chirp mass $\mathcal{M}_{z}$, we fix $\mathcal{M}_{z1}$ and vary $\mathcal{M}_{z2}$. In particular,  in Figs.~\ref{fig:Contour_Error_BNS} and \ref{fig:Contour_Error_BBH}, we rewrite $\mathcal{M}_{z2}$ with $\eta_2^{3/5}M_{z2}$ and vary the symmetric mass ratio $\eta_2$, fixing $M_{z2}=M_{z1}$. One may note that the results presented here actually depend on how we vary the parameters $t_{\rm c2}-t_{\rm c1}$ and $\mathcal{M}_{z2}-\mathcal{M}_{z1}$. However, the resulting behaviors are essentially the same except for a detailed structure of the statistical errors or systematic biases, and the conclusion of this paper remains unchanged irrespective of the choice of the variation of parameters.

Figure~\ref{fig:Contour_Error_BNS} plots the expected errors on the parameters of the overlapping BNS-BNS$^\prime$ events, while Fig.~\ref{fig:Contour_Error_BBH} shows those of the overlapping BBH1-BBH1$^\prime$ case. In each panel, the estimated error on each parameter is normalized by the one in the single-event case, whose value is indicated in each panel (see also Table \ref{tab:single_event_errors}). The results at the 2PN order are shown in the lower panels. For comparison, we also plot the Newtonian results in the upper panels, together with the critical conditions discussed in Sec.~\ref{subsec:analytical} (depicted as dotted lines). Looking at the overlapping BNS-BNS$^\prime$ cases, the statistical errors on the three parameters are found to remain almost the same as those obtained from the single-event case. Even in extreme cases with $|\mathcal{M}_{z2}-\mathcal{M}_{z1}|\sim 10^{-4}\,M_{\odot}$, the errors get slightly worse and amount to $15\%-20\%$ at $|t_{\rm c2}-t_{\rm c1}|\sim10^{-3}\,{\rm s}$. The main reason why no significant degradation occurs in the overlapping BNS-BNS$^\prime$ cases is that for an ET-like detector, the total signal duration is sufficiently longer than the phase-matching signal duration, for which the signal frequencies at a time almost coincide. By contrast, in Fig.~\ref{fig:Contour_Error_BBH} the observation time of the BBH systems is much shorter than that of BNS, and thus the overlapping BBH1-BBH$^\prime$ cases exhibit a large error on each parameter. Nevertheless, a significant degradation of the error appears only at the region of 
$|\mathcal{M}_{z2}-\mathcal{M}_{z1}|\lesssim 10^{-1}\,M_{\odot}$ and 
$|t_{\rm c2}-t_{\rm c1}|\lesssim 10^{-1}\,{\rm s}$, where the occurrence of overlapping events is rather small, well below the white stripe, meaning that the number of overlapping signals producing a large error is expected to be very small. In any case, the analytically estimated critical conditions derived at the Newtonian order provide a useful guideline and explain the region where we see a large error at the 2PN order.

In Fig.~\ref{fig:Contour_Bias}, as another representative case we consider the overlapping BBH2-BBH3 events and estimate the systematic biases in the best-fit values of the parameters $t_{\rm c}$ (left panel), $\phi_{\rm c}$ (middle panle), and $\mathcal{M}_z$ (right panel). Note that for BNS cases, the phase-matching signal duration, for which the signal frequencies at a time almost  coincide, is rather short, and no significant impact is expected for the biases in the parameter estimation. In Fig.~\ref{fig:Contour_Bias}, apart from the BBH3 system having a lower SNR, the overlapping signal is misinterpreted as a single GW event coming from the BBH2 system, and the biases in the estimated parameters of the BBH2 system are evaluated based on the formalism in Sec.~\ref{subsubsec:systematic_bias}. With the coalescence time and redshifted chirp mass of the BBH3 system slightly varied, the resultant biases are plotted as a function of differences of the coalescence times and redshifted chirp masses. The estimated values of the biases at the 2PN order are then normalized by the expected 1$\sigma$ errors in the single-event case. Over the plotted region, the resultant biases typically have $10\%-30\%$  variation in amplitude relative to the statistical error in the single-event case. A closer look at the region depicted in light blue reveals an oscillatory behavior, although we cannot resolve such a fine structure with the color scale adopted in Fig.~\ref{fig:Contour_Bias}. Note that sometimes a large bias exceeding more than $\sim50\%$ accidentally happens (pixels in dark red or blue), but such a case is restricted to a very specific parameter region. Hence, in the majority of the parameter regions the systematic biases are insignificant, and we conclude that there is no serious impact on the parameter estimation.

Although this conclusion strictly holds at the 2PN order, we have also examined the cases at the Newtonian, 1, and 1.5PN orders. As illustrated in Table~\ref{tab:single_event_errors}, 
if we increase the PN order, the number of parameters to estimate increases, and the statistical error on each parameter typically becomes large. On the other hand, the size of the {\it unnormalized} biases in the parameter estimation, i.e., $\Delta t_{\rm c}$, $\Delta\phi_{\rm c}$, and $\Delta\mathcal{M}_{z}$, does not change much. Instead, we found that it is insensitive to the PN order. This suggests that, even at the higher PN order, the bias in the best-fit value remains insignificant with respect to the statistical error. We thus expect that the same conclusion obtained at the 2PN order generally holds. This would be the case even if we consider nonzero spins since they give additional contributions to the GW phases at $1.5$PN, thereby reducing the overlap of GW signals.


\section{Conclusion}
\label{sec:conclusion}

The gravitational-wave (GW) observations via third-generation laser interferometers will open up various new windows to study and understand the entire history of the Universe. In particular, with an improved detector sensitivity, a dramatic increase in the number of GW events is expected,  and the number of detectable redshifts of the GW sources will become increasingly high. This will, however, present several unexpected issues that have not previously been explored. The impact of the overlapping GW events is one such issue.

Improving the detection efficiency, we will be able to detect the two different GW signals,  coming from the binary black holes (BBHs) and binary neutron stars (BNSs), that eventually overlap with each other. The expected number of such an event is thought to be non-negligible, and their impacts on the parameter estimation study might be significant. To be precise, in the presence of the overlapping GW signals, one cannot separately treat these two signals in the parameter estimation analysis, and the parameter degeneracy between these overlapping systems can happen, potentially leading to a substantial error on the parameter of each system. Furthermore, the signal-to-noise ratios for each of the overlapping binaries would not be the same, and the overlapping events may consist of a pair of loud and quiet binaries. In such a case, we may misinterpret the event as a single GW signal, and this can potentially lead to a serious bias in the estimated parameters. Since the occurrence of substantial errors or systematic biases in the parameter estimation yields an imperfect foreground noise subtraction,  the overlapping GW events might also give a large impact on the detection of stochastic GW backgrounds.

In this paper, based on the Monte Carlo simulations, we first estimated the expected number of overlapping GW signals detectable with a third-generation detector like the Einstein Telescope (ET). We found that there are non-negligible pairs of binary systems detected during a one-year observation whose coalescence times are very close to each other,
$|t_{\rm c1}-t_{\rm c2}|\lesssim10^{-2}$\,s for BNSs and $1$\,s for BBHs (see Fig.~\ref{fig:tc-diff-dist}). 
Among these overlapping events, there are furthermore a certain amount of detectable events having comparable redshifted chirp masses, $\mathcal{M}_{z1}\simeq\mathcal{M}_{z2}$ (see Fig.~\ref{fig:tc-mc-diff-dist}). These events potentially lead to a large error or biased parameter estimation. We then used the Fisher matrix formalism to quantitatively investigate the size of the statistical errors on the parameters of representative binary systems.

Our finding is that the overlapping signals do not produce large statistical errors on the parameters of each binary system unless the coalescence times and the redshifted chirp masses of the two overlapping GWs are very close to each other. The occurrence rate of such a closely overlapping event is rather small even with an ET-like detector. We also estimated a possible bias in the best-fit parameters of the overlapping BBH systems. We found that erroneously analyzing the overlapping GWs as a single GW event, the best-fit parameters are prone to be biased typically at a level of $10\% - 30\%$ relative to the statistical errors computed from the single-event case. Therefore, the overlapping binary signals detectable with ET-like third-generation laser interferometers do not have a serious impact on the parameter estimation of the binary systems. Strictly, this conclusion is valid only for the analysis at the second post-Newtonian (PN) order, but the trends and behaviors at different PN orders suggests that the conclusion holds even at the higher PN orders.

Our conclusion readily implies that, except for the rare cases in which the parameters are very close to each other, most of the overlapping two GW signals in the time domain are safely distinguishable using the matched filtering method, and hence 
the residual noise of the GWs after subtracting the astrophysical foregrounds may not be a serious issue in detecting the stochastic backgrounds of the cosmological origin via third-generation detectors. Instead, the detectability of the stochastic backgrounds would be severely limited by the confusion noise formed by numerous GW events with a small SNR. Another potential concern is the correlated noise induced by the global magnetic fields in the Earth-ionosphere cavity, known as the Schumann resonance. It could prevent us from gaining a solid confirmation of the stochastic GWs, especially at low-frequency bands (see, e.g., \cite{2013PhRvD..87l3009T, 2014PhRvD..90b3013T, 2017PhRvD..96b2004H, 2019PhRvD.100h2001H, 2020PhRvD.102j2005M}). Toward the search for cosmological backgrounds, development of methodologies to efficiently subtract low-SNR GW events as well as to mitigate the correlated magnetic noise is crucial. We leave this for future work.

\begin{acknowledgments}
This work was supported in part by MEXT/JSPS KAKENHI Grants No. JP21K03580 (Y.H.), Grants No. JP17H06359, No. JP20H05861, and No. JP21H01081 (A.T.), and Grants No. JP19H01894 and JP20H04726 (A.N.). 
A.T. acknowledges the support from JST AIP Acceleration Research Grant No. JP20317829, Japan. A.N. was also supported by research grants from the Inamori Foundation.
\end{acknowledgments}

\appendix
\section{Derivation of Eq.~(\ref{eq:formula_systematic_bias})}
\label{sec:derivation}

In this Appendix, we derive the analytical formula to quantify the systematic bias given in Eq.~(\ref{eq:formula_systematic_bias}). Consider the GW signal made of the two overlapping merger events, 1 and 2. We denote it by $s=s_{1}$ and $s_{2}$. In estimating the GW parameters, we are particularly concerned with the case in which the signal is misinterpreted as a single GW event, and we ignore the event 2. That is, instead of the overlapping template of $h=h_{1} + h_{2}$, the wrong template with a single GW event, $h=h_{1}$, is adopted, and the best-fit parameters for the event 1 are derived by maximizing the likelihood function, $\mathcal{L}\propto \exp\{-\chi^2/2\}$, with the function $\chi^2$ given by
\begin{align}
    \chi^2\equiv (s-h|s-h).
    \label{eq:chi_function}
\end{align}
Here, the inner product $(A|B)$ is defined as 
\begin{align}
(A|B)   \equiv 2\int \frac{ \tilde{A}^*(f) \tilde{B}(f) +\tilde{A}(f) \tilde{B}^*(f) }{S_{\rm{n}}(f)}\,df.
\end{align}

In order to explicitly see the impact of using the wrong template $h=h_{1}$ on the best-fit parameters, one may decompose the template $h$ into two pieces,
\begin{align}
    h&=h^{\rm true}+h^{\rm sys}
\end{align}
with the templates $h^{\rm true}$ and $h^{\rm sys}$, respectively, corresponding to a proper template for the overlapping signal and the one ignored in the likelihood analysis-and hence regarded as systematics-given by
\begin{align}
    h^{\rm true}= h_{1}+h_{2},\quad h^{\rm sys}=-h_{2}.
\end{align}
Equation~(\ref{eq:chi_function}) is then rewritten as follows:
\begin{align}
    \chi^2&=(h^{\rm true}-s|h^{\rm true}-s)+(h^{\rm sys}|h^{\rm sys})
    \nonumber\\
    &+2(h^{\rm sys}|h^{\rm true}-s)
    \label{eq:chi2_true_sys}
\end{align}
The best-fit parameter for event 1, $\theta^{{\rm best}}_{1 a}$, is obtained by extremizing the function $\chi^2$. If one adopts a proper template, the best-fit values reproduce the parameters for a fiducial setup, $\theta_{a}^{\rm fid}$, but, due to the systematics in the template, the best-fit values can deviate from the fiducial values. One has
\begin{align}
    0&=\left.\frac{\partial \chi^2}{\partial \theta_{1 a}}\right|_{\mbox{\boldmath$\theta$}_{1}^{\rm best}}
 \nonumber
\\   
&\simeq\left.\frac{\partial \chi^2}{\partial \theta_{1 a}}\right|_{\mbox{\boldmath$\theta$}^{\rm fid}_{1}}+\sum_b \left.\frac{\partial^2 \chi^2}{\partial \theta_{1 a} \partial \theta_{1 b}}\right|_{\mbox{\boldmath$\theta$}^{\rm fid}_{1}} \Delta\theta_{1 b},
\label{eq:extremum_chi2}
\end{align}
where, on the second line, we have expanded the derivative of the $\chi^2$ function around the fiducial parameters, and assumed that the difference between the best-fit and fiducial values, defined as $\Delta\theta_{1 b}\equiv \theta_{1 b}^{\rm best}-\theta_{1 b}^{\rm fid}$, is small. 

Using the expression in Eq.~(\ref{eq:chi2_true_sys}), we compute the derivatives of the function $\chi^2$ in Eq.~(\ref{eq:extremum_chi2}) as follows:
\begin{align}
    \left.\frac{\partial \chi^2}{\partial \theta_{1 a}}\right|_{\mbox{\boldmath$\theta$}^{\rm fid}_{1}}&=2\Bigl(h^{\rm sys}\Bigl|\frac{\partial h_{1}}{\partial \theta_{1 a}}\Bigr),
\\
    \left.\frac{\partial^2 \chi^2}{\partial \theta_{1 a} \partial \theta_{1 b}}\right|_{\mbox{\boldmath$\theta$}^{\rm fid}_{1} }&=
    2\Bigl(\frac{\partial h_{1}}{\partial \theta_{1 a}}\Bigl|\frac{\partial h_{1}}{\partial \theta_{1 b}}\Bigr)+
    2\Bigl(h^{\rm sys}\Bigl|\frac{\partial^2 h_{1}}{\partial \theta_{1 a}\partial \theta_{1 b}}\Bigr),
\end{align}
where we use the fact that the signal $s$ is identical to the template $h^{\rm true}$ when we evaluate it with the fiducial parameters.

With these expressions, Eq.~(\ref{eq:extremum_chi2}) can be recast as
\begin{align}
&    0=- s_a +\sum_b {\mathcal F}_{ab}\Delta\theta_{1 b}\,
\end{align}
with the vector $s_a$ and matrix $\mathcal{F}_{ab}$ given by 
\begin{align}
s_a&= \Bigl(h_{2}\Bigl|\frac{\partial h_{1}}{\partial \theta_{1 a}}\Bigr),
\\
{\mathcal F}_{ab}&=
    \Bigl(\frac{\partial h_{1}}{\partial \theta_{1 a}}\Bigl|\frac{\partial h_{1}}{\partial \theta_{1 b}}\Bigr)-
    \Bigl(h_{2}\Bigl|\frac{\partial^2 h_{1}}{\partial \theta_{1 a}\partial \theta_{1 b}}\Bigr)
\end{align}
where we use $h^{\rm sys}=-h_{2}$. These are the exact same vector and matrix quantities as those given in Eqs.~(\ref{eq:vector_s}) and (\ref{eq:matrix_F}), respectively. Hence, solving the above equation with respect to $\Delta\theta_{1 b}$ immediately leads to Eq.~(\ref{eq:formula_systematic_bias}).


\begin{thebibliography}{33}%
\makeatletter
\providecommand \@ifxundefined [1]{%
 \@ifx{#1\undefined}
}%
\providecommand \@ifnum [1]{%
 \ifnum #1\expandafter \@firstoftwo
 \else \expandafter \@secondoftwo
 \fi
}%
\providecommand \@ifx [1]{%
 \ifx #1\expandafter \@firstoftwo
 \else \expandafter \@secondoftwo
 \fi
}%
\providecommand \natexlab [1]{#1}%
\providecommand \enquote  [1]{``#1''}%
\providecommand \bibnamefont  [1]{#1}%
\providecommand \bibfnamefont [1]{#1}%
\providecommand \citenamefont [1]{#1}%
\providecommand \href@noop [0]{\@secondoftwo}%
\providecommand \href [0]{\begingroup \@sanitize@url \@href}%
\providecommand \@href[1]{\@@startlink{#1}\@@href}%
\providecommand \@@href[1]{\endgroup#1\@@endlink}%
\providecommand \@sanitize@url [0]{\catcode `\\12\catcode `\$12\catcode
  `\&12\catcode `\#12\catcode `\^12\catcode `\_12\catcode `\%12\relax}%
\providecommand \@@startlink[1]{}%
\providecommand \@@endlink[0]{}%
\providecommand \url  [0]{\begingroup\@sanitize@url \@url }%
\providecommand \@url [1]{\endgroup\@href {#1}{\urlprefix }}%
\providecommand \urlprefix  [0]{URL }%
\providecommand \Eprint [0]{\href }%
\providecommand \doibase [0]{http://dx.doi.org/}%
\providecommand \selectlanguage [0]{\@gobble}%
\providecommand \bibinfo  [0]{\@secondoftwo}%
\providecommand \bibfield  [0]{\@secondoftwo}%
\providecommand \translation [1]{[#1]}%
\providecommand \BibitemOpen [0]{}%
\providecommand \bibitemStop [0]{}%
\providecommand \bibitemNoStop [0]{.\EOS\space}%
\providecommand \EOS [0]{\spacefactor3000\relax}%
\providecommand \BibitemShut  [1]{\csname bibitem#1\endcsname}%
\let\auto@bib@innerbib\@empty
\bibitem [{\citenamefont {Aasi}\ \emph {et~al.}(2015)\citenamefont {Aasi} \emph
  {et~al.}}]{TheLIGOScientific:2014jea}%
  \BibitemOpen
  \bibfield  {author} {\bibinfo {author} {\bibfnamefont {J.}~\bibnamefont
  {Aasi}} \emph {et~al.} (\bibinfo {collaboration} {LIGO Scientific}),\ }\href
  {\doibase 10.1088/0264-9381/32/7/074001} {\bibfield  {journal} {\bibinfo
  {journal} {Class. Quant. Grav.}\ }\textbf {\bibinfo {volume} {32}},\ \bibinfo
  {pages} {074001} (\bibinfo {year} {2015})},\ \Eprint
  {http://arxiv.org/abs/1411.4547} {arXiv:1411.4547 [gr-qc]} \BibitemShut
  {NoStop}%
\bibitem [{\citenamefont {Acernese}\ \emph {et~al.}(2015)\citenamefont
  {Acernese} \emph {et~al.}}]{TheVirgo:2014hva}%
  \BibitemOpen
  \bibfield  {author} {\bibinfo {author} {\bibfnamefont {F.}~\bibnamefont
  {Acernese}} \emph {et~al.} (\bibinfo {collaboration} {VIRGO}),\ }\href
  {\doibase 10.1088/0264-9381/32/2/024001} {\bibfield  {journal} {\bibinfo
  {journal} {Class. Quant. Grav.}\ }\textbf {\bibinfo {volume} {32}},\ \bibinfo
  {pages} {024001} (\bibinfo {year} {2015})},\ \Eprint
  {http://arxiv.org/abs/1408.3978} {arXiv:1408.3978 [gr-qc]} \BibitemShut
  {NoStop}%
\bibitem [{\citenamefont {Abbott}\ \emph {et~al.}(2019)\citenamefont {Abbott}
  \emph {et~al.}}]{LIGOScientific:2018mvr}%
  \BibitemOpen
  \bibfield  {author} {\bibinfo {author} {\bibfnamefont {B.}~\bibnamefont
  {Abbott}} \emph {et~al.} (\bibinfo {collaboration} {LIGO Scientific,
  Virgo}),\ }\href {\doibase 10.1103/PhysRevX.9.031040} {\bibfield  {journal}
  {\bibinfo  {journal} {Phys. Rev. X}\ }\textbf {\bibinfo {volume} {9}},\
  \bibinfo {pages} {031040} (\bibinfo {year} {2019})},\ \Eprint
  {http://arxiv.org/abs/1811.12907} {arXiv:1811.12907 [astro-ph.HE]}
  \BibitemShut {NoStop}%
\bibitem [{\citenamefont {Abbott}\ \emph
  {et~al.}(2020{\natexlab{a}})\citenamefont {Abbott} \emph
  {et~al.}}]{Abbott:2020niy}%
  \BibitemOpen
  \bibfield  {author} {\bibinfo {author} {\bibfnamefont {R.}~\bibnamefont
  {Abbott}} \emph {et~al.} (\bibinfo {collaboration} {LIGO Scientific,
  Virgo}),\ }\href@noop {} {\  (\bibinfo {year} {2020}{\natexlab{a}})},\
  \Eprint {http://arxiv.org/abs/2010.14527} {arXiv:2010.14527 [gr-qc]}
  \BibitemShut {NoStop}%
\bibitem [{ET:()}]{ET:2020}%
  \BibitemOpen
  \href@noop {} {}\bibinfo {note} {Design Report Update 2020 for the Einstein
  Telescope: https://apps.et-gw.eu/tds/?content=3\&r=17245}\BibitemShut
  {NoStop}%
\bibitem [{\citenamefont {Abbott}\ \emph {et~al.}(2017)\citenamefont {Abbott}
  \emph {et~al.}}]{Evans:2016mbw}%
  \BibitemOpen
  \bibfield  {author} {\bibinfo {author} {\bibfnamefont {B.~P.}\ \bibnamefont
  {Abbott}} \emph {et~al.} (\bibinfo {collaboration} {LIGO Scientific}),\
  }\href {\doibase 10.1088/1361-6382/aa51f4} {\bibfield  {journal} {\bibinfo
  {journal} {Class. Quant. Grav.}\ }\textbf {\bibinfo {volume} {34}},\ \bibinfo
  {pages} {044001} (\bibinfo {year} {2017})},\ \Eprint
  {http://arxiv.org/abs/1607.08697} {arXiv:1607.08697 [astro-ph.IM]}
  \BibitemShut {NoStop}%
\bibitem [{\citenamefont {Nishizawa}(2016)}]{Nishizawa:2016kba}%
  \BibitemOpen
  \bibfield  {author} {\bibinfo {author} {\bibfnamefont {A.}~\bibnamefont
  {Nishizawa}},\ }\href {\doibase 10.1103/PhysRevD.93.124036} {\bibfield
  {journal} {\bibinfo  {journal} {Phys. Rev.}\ }\textbf {\bibinfo {volume}
  {D93}},\ \bibinfo {pages} {124036} (\bibinfo {year} {2016})},\ \Eprint
  {http://arxiv.org/abs/1601.01072} {arXiv:1601.01072 [gr-qc]} \BibitemShut
  {NoStop}%
\bibitem [{\citenamefont {Samajdar}\ \emph {et~al.}(2021)\citenamefont
  {Samajdar}, \citenamefont {Janquart}, \citenamefont {Van Den~Broeck},\ and\
  \citenamefont {Dietrich}}]{Samajdar:2021egv}%
  \BibitemOpen
  \bibfield  {author} {\bibinfo {author} {\bibfnamefont {A.}~\bibnamefont
  {Samajdar}}, \bibinfo {author} {\bibfnamefont {J.}~\bibnamefont {Janquart}},
  \bibinfo {author} {\bibfnamefont {C.}~\bibnamefont {Van Den~Broeck}}, \ and\
  \bibinfo {author} {\bibfnamefont {T.}~\bibnamefont {Dietrich}},\ }\href@noop
  {} {\  (\bibinfo {year} {2021})},\ \Eprint {http://arxiv.org/abs/2102.07544}
  {arXiv:2102.07544 [gr-qc]} \BibitemShut {NoStop}%
\bibitem [{\citenamefont {Pizzati}\ \emph {et~al.}(2021)\citenamefont
  {Pizzati}, \citenamefont {Sachdev}, \citenamefont {Gupta},\ and\
  \citenamefont {Sathyaprakash}}]{Pizzati:2021gzd}%
  \BibitemOpen
  \bibfield  {author} {\bibinfo {author} {\bibfnamefont {E.}~\bibnamefont
  {Pizzati}}, \bibinfo {author} {\bibfnamefont {S.}~\bibnamefont {Sachdev}},
  \bibinfo {author} {\bibfnamefont {A.}~\bibnamefont {Gupta}}, \ and\ \bibinfo
  {author} {\bibfnamefont {B.}~\bibnamefont {Sathyaprakash}},\ }\href@noop {}
  {\  (\bibinfo {year} {2021})},\ \Eprint {http://arxiv.org/abs/2102.07692}
  {arXiv:2102.07692 [gr-qc]} \BibitemShut {NoStop}%
\bibitem [{\citenamefont {Maggiore}(2000)}]{Maggiore:1999vm}%
  \BibitemOpen
  \bibfield  {author} {\bibinfo {author} {\bibfnamefont {M.}~\bibnamefont
  {Maggiore}},\ }\href {\doibase 10.1016/S0370-1573(99)00102-7} {\bibfield
  {journal} {\bibinfo  {journal} {Phys.Rept.}\ }\textbf {\bibinfo {volume}
  {331}},\ \bibinfo {pages} {283} (\bibinfo {year} {2000})},\ \Eprint
  {http://arxiv.org/abs/gr-qc/9909001} {arXiv:gr-qc/9909001 [gr-qc]}
  \BibitemShut {NoStop}%
\bibitem [{\citenamefont {{Caprini}}\ and\ \citenamefont
  {{Figueroa}}(2018)}]{2018CQGra..35p3001C}%
  \BibitemOpen
  \bibfield  {author} {\bibinfo {author} {\bibfnamefont {C.}~\bibnamefont
  {{Caprini}}}\ and\ \bibinfo {author} {\bibfnamefont {D.~G.}\ \bibnamefont
  {{Figueroa}}},\ }\href {\doibase 10.1088/1361-6382/aac608} {\bibfield
  {journal} {\bibinfo  {journal} {Classical and Quantum Gravity}\ }\textbf
  {\bibinfo {volume} {35}},\ \bibinfo {eid} {163001} (\bibinfo {year}
  {2018})},\ \Eprint {http://arxiv.org/abs/1801.04268} {arXiv:1801.04268
  [astro-ph.CO]} \BibitemShut {NoStop}%
\bibitem [{\citenamefont {{Cutler}}\ and\ \citenamefont
  {{Harms}}(2006)}]{Cutler:2006}%
  \BibitemOpen
  \bibfield  {author} {\bibinfo {author} {\bibfnamefont {C.}~\bibnamefont
  {{Cutler}}}\ and\ \bibinfo {author} {\bibfnamefont {J.}~\bibnamefont
  {{Harms}}},\ }\href {\doibase 10.1103/PhysRevD.73.042001} {\bibfield
  {journal} {\bibinfo  {journal} {Phys. Rev. D}\ }\textbf {\bibinfo {volume}
  {73}},\ \bibinfo {eid} {042001} (\bibinfo {year} {2006})},\ \Eprint
  {http://arxiv.org/abs/gr-qc/0511092} {gr-qc/0511092} \BibitemShut {NoStop}%
\bibitem [{\citenamefont {Harms}\ \emph {et~al.}(2008)\citenamefont {Harms},
  \citenamefont {Mahrdt}, \citenamefont {Otto},\ and\ \citenamefont
  {Priess}}]{Harms:2008xv}%
  \BibitemOpen
  \bibfield  {author} {\bibinfo {author} {\bibfnamefont {J.}~\bibnamefont
  {Harms}}, \bibinfo {author} {\bibfnamefont {C.}~\bibnamefont {Mahrdt}},
  \bibinfo {author} {\bibfnamefont {M.}~\bibnamefont {Otto}}, \ and\ \bibinfo
  {author} {\bibfnamefont {M.}~\bibnamefont {Priess}},\ }\href {\doibase
  10.1103/PhysRevD.77.123010} {\bibfield  {journal} {\bibinfo  {journal} {Phys.
  Rev. D}\ }\textbf {\bibinfo {volume} {77}},\ \bibinfo {pages} {123010}
  (\bibinfo {year} {2008})},\ \Eprint {http://arxiv.org/abs/0803.0226}
  {arXiv:0803.0226 [gr-qc]} \BibitemShut {NoStop}%
\bibitem [{\citenamefont {Yagi}\ and\ \citenamefont
  {Seto}(2011)}]{Yagi:2011wg}%
  \BibitemOpen
  \bibfield  {author} {\bibinfo {author} {\bibfnamefont {K.}~\bibnamefont
  {Yagi}}\ and\ \bibinfo {author} {\bibfnamefont {N.}~\bibnamefont {Seto}},\
  }\href {\doibase 10.1103/PhysRevD.83.044011} {\bibfield  {journal} {\bibinfo
  {journal} {Phys. Rev. D}\ }\textbf {\bibinfo {volume} {83}},\ \bibinfo
  {pages} {044011} (\bibinfo {year} {2011})},\ \bibinfo {note} {[Erratum:
  Phys.Rev.D 95, 109901 (2017)]},\ \Eprint {http://arxiv.org/abs/1101.3940}
  {arXiv:1101.3940 [astro-ph.CO]} \BibitemShut {NoStop}%
\bibitem [{\citenamefont {Nishizawa}\ \emph {et~al.}(2012)\citenamefont
  {Nishizawa}, \citenamefont {Yagi}, \citenamefont {Taruya},\ and\
  \citenamefont {Tanaka}}]{Nishizawa:2011eq}%
  \BibitemOpen
  \bibfield  {author} {\bibinfo {author} {\bibfnamefont {A.}~\bibnamefont
  {Nishizawa}}, \bibinfo {author} {\bibfnamefont {K.}~\bibnamefont {Yagi}},
  \bibinfo {author} {\bibfnamefont {A.}~\bibnamefont {Taruya}}, \ and\ \bibinfo
  {author} {\bibfnamefont {T.}~\bibnamefont {Tanaka}},\ }\href {\doibase
  10.1103/PhysRevD.85.044047} {\bibfield  {journal} {\bibinfo  {journal} {Phys.
  Rev.}\ }\textbf {\bibinfo {volume} {D85}},\ \bibinfo {pages} {044047}
  (\bibinfo {year} {2012})},\ \Eprint {http://arxiv.org/abs/1110.2865}
  {arXiv:1110.2865 [astro-ph.CO]} \BibitemShut {NoStop}%
\bibitem [{\citenamefont {Adams}\ and\ \citenamefont
  {Cornish}(2014)}]{Adams:2013qma}%
  \BibitemOpen
  \bibfield  {author} {\bibinfo {author} {\bibfnamefont {M.~R.}\ \bibnamefont
  {Adams}}\ and\ \bibinfo {author} {\bibfnamefont {N.~J.}\ \bibnamefont
  {Cornish}},\ }\href {\doibase 10.1103/PhysRevD.89.022001} {\bibfield
  {journal} {\bibinfo  {journal} {Phys. Rev. D}\ }\textbf {\bibinfo {volume}
  {89}},\ \bibinfo {pages} {022001} (\bibinfo {year} {2014})},\ \Eprint
  {http://arxiv.org/abs/1307.4116} {arXiv:1307.4116 [gr-qc]} \BibitemShut
  {NoStop}%
\bibitem [{\citenamefont {Seto}(2009)}]{Seto:2009bf}%
  \BibitemOpen
  \bibfield  {author} {\bibinfo {author} {\bibfnamefont {N.}~\bibnamefont
  {Seto}},\ }\href {\doibase 10.1103/PhysRevD.80.103001} {\bibfield  {journal}
  {\bibinfo  {journal} {Phys. Rev. D}\ }\textbf {\bibinfo {volume} {80}},\
  \bibinfo {pages} {103001} (\bibinfo {year} {2009})},\ \Eprint
  {http://arxiv.org/abs/0910.4812} {arXiv:0910.4812 [gr-qc]} \BibitemShut
  {NoStop}%
\bibitem [{\citenamefont {Regimbau}\ \emph {et~al.}(2017)\citenamefont
  {Regimbau}, \citenamefont {Evans}, \citenamefont {Christensen}, \citenamefont
  {Katsavounidis}, \citenamefont {Sathyaprakash},\ and\ \citenamefont
  {Vitale}}]{Regimbau:2016ike}%
  \BibitemOpen
  \bibfield  {author} {\bibinfo {author} {\bibfnamefont {T.}~\bibnamefont
  {Regimbau}}, \bibinfo {author} {\bibfnamefont {M.}~\bibnamefont {Evans}},
  \bibinfo {author} {\bibfnamefont {N.}~\bibnamefont {Christensen}}, \bibinfo
  {author} {\bibfnamefont {E.}~\bibnamefont {Katsavounidis}}, \bibinfo {author}
  {\bibfnamefont {B.}~\bibnamefont {Sathyaprakash}}, \ and\ \bibinfo {author}
  {\bibfnamefont {S.}~\bibnamefont {Vitale}},\ }\href {\doibase
  10.1103/PhysRevLett.118.151105} {\bibfield  {journal} {\bibinfo  {journal}
  {Phys. Rev. Lett.}\ }\textbf {\bibinfo {volume} {118}},\ \bibinfo {pages}
  {151105} (\bibinfo {year} {2017})},\ \Eprint
  {http://arxiv.org/abs/1611.08943} {arXiv:1611.08943 [astro-ph.CO]}
  \BibitemShut {NoStop}%
\bibitem [{\citenamefont {Sachdev}\ \emph {et~al.}(2020)\citenamefont
  {Sachdev}, \citenamefont {Regimbau},\ and\ \citenamefont
  {Sathyaprakash}}]{Sachdev:2020bkk}%
  \BibitemOpen
  \bibfield  {author} {\bibinfo {author} {\bibfnamefont {S.}~\bibnamefont
  {Sachdev}}, \bibinfo {author} {\bibfnamefont {T.}~\bibnamefont {Regimbau}}, \
  and\ \bibinfo {author} {\bibfnamefont {B.}~\bibnamefont {Sathyaprakash}},\
  }\href {\doibase 10.1103/PhysRevD.102.024051} {\bibfield  {journal} {\bibinfo
   {journal} {Phys. Rev. D}\ }\textbf {\bibinfo {volume} {102}},\ \bibinfo
  {pages} {024051} (\bibinfo {year} {2020})},\ \Eprint
  {http://arxiv.org/abs/2002.05365} {arXiv:2002.05365 [gr-qc]} \BibitemShut
  {NoStop}%
\bibitem [{\citenamefont {Sharma}\ and\ \citenamefont
  {Harms}(2020)}]{Sharma:2020btq}%
  \BibitemOpen
  \bibfield  {author} {\bibinfo {author} {\bibfnamefont {A.}~\bibnamefont
  {Sharma}}\ and\ \bibinfo {author} {\bibfnamefont {J.}~\bibnamefont {Harms}},\
  }\href {\doibase 10.1103/PhysRevD.102.063009} {\bibfield  {journal} {\bibinfo
   {journal} {Phys. Rev. D}\ }\textbf {\bibinfo {volume} {102}},\ \bibinfo
  {pages} {063009} (\bibinfo {year} {2020})},\ \Eprint
  {http://arxiv.org/abs/2006.16116} {arXiv:2006.16116 [gr-qc]} \BibitemShut
  {NoStop}%
\bibitem [{\citenamefont {Crowder}\ and\ \citenamefont
  {Cornish}(2004)}]{Crowder:2004ca}%
  \BibitemOpen
  \bibfield  {author} {\bibinfo {author} {\bibfnamefont {J.}~\bibnamefont
  {Crowder}}\ and\ \bibinfo {author} {\bibfnamefont {N.~J.}\ \bibnamefont
  {Cornish}},\ }\href {\doibase 10.1103/PhysRevD.70.082004} {\bibfield
  {journal} {\bibinfo  {journal} {Phys. Rev. D}\ }\textbf {\bibinfo {volume}
  {70}},\ \bibinfo {pages} {082004} (\bibinfo {year} {2004})},\ \Eprint
  {http://arxiv.org/abs/gr-qc/0404129} {arXiv:gr-qc/0404129} \BibitemShut
  {NoStop}%
\bibitem [{\citenamefont {Nishizawa}(2017)}]{Nishizawa:2016ood}%
  \BibitemOpen
  \bibfield  {author} {\bibinfo {author} {\bibfnamefont {A.}~\bibnamefont
  {Nishizawa}},\ }\href {\doibase 10.1103/PhysRevD.96.101303} {\bibfield
  {journal} {\bibinfo  {journal} {Phys. Rev.}\ }\textbf {\bibinfo {volume}
  {D96}},\ \bibinfo {pages} {101303} (\bibinfo {year} {2017})},\ \Eprint
  {http://arxiv.org/abs/1612.06060} {arXiv:1612.06060 [astro-ph.CO]}
  \BibitemShut {NoStop}%
\bibitem [{\citenamefont {Aghanim}\ \emph {et~al.}(2018)\citenamefont {Aghanim}
  \emph {et~al.}}]{Planck2018cosmology}%
  \BibitemOpen
  \bibfield  {author} {\bibinfo {author} {\bibfnamefont {N.}~\bibnamefont
  {Aghanim}} \emph {et~al.} (\bibinfo {collaboration} {Planck}),\ }\href@noop
  {} {\  (\bibinfo {year} {2018})},\ \Eprint {http://arxiv.org/abs/1807.06209}
  {arXiv:1807.06209 [astro-ph.CO]} \BibitemShut {NoStop}%
\bibitem [{\citenamefont {Abbott}\ \emph
  {et~al.}(2020{\natexlab{b}})\citenamefont {Abbott} \emph
  {et~al.}}]{Abbott:2020gyp}%
  \BibitemOpen
  \bibfield  {author} {\bibinfo {author} {\bibfnamefont {R.}~\bibnamefont
  {Abbott}} \emph {et~al.} (\bibinfo {collaboration} {LIGO Scientific,
  Virgo}),\ }\href@noop {} {\  (\bibinfo {year} {2020}{\natexlab{b}})},\
  \Eprint {http://arxiv.org/abs/2010.14533} {arXiv:2010.14533 [astro-ph.HE]}
  \BibitemShut {NoStop}%
\bibitem [{\citenamefont {Nishizawa}\ and\ \citenamefont
  {Arai}(2019)}]{Nishizawa:2019rra}%
  \BibitemOpen
  \bibfield  {author} {\bibinfo {author} {\bibfnamefont {A.}~\bibnamefont
  {Nishizawa}}\ and\ \bibinfo {author} {\bibfnamefont {S.}~\bibnamefont
  {Arai}},\ }\href {\doibase 10.1103/PhysRevD.99.104038} {\bibfield  {journal}
  {\bibinfo  {journal} {Phys. Rev.}\ }\textbf {\bibinfo {volume} {D99}},\
  \bibinfo {pages} {104038} (\bibinfo {year} {2019})},\ \Eprint
  {http://arxiv.org/abs/1901.08249} {arXiv:1901.08249 [gr-qc]} \BibitemShut
  {NoStop}%
\bibitem [{\citenamefont {Cutler}\ and\ \citenamefont
  {Flanagan}(1994)}]{Cutler:1994ys}%
  \BibitemOpen
  \bibfield  {author} {\bibinfo {author} {\bibfnamefont {C.}~\bibnamefont
  {Cutler}}\ and\ \bibinfo {author} {\bibfnamefont {E.~E.}\ \bibnamefont
  {Flanagan}},\ }\href {\doibase 10.1103/PhysRevD.49.2658} {\bibfield
  {journal} {\bibinfo  {journal} {Phys.Rev.}\ }\textbf {\bibinfo {volume}
  {D49}},\ \bibinfo {pages} {2658} (\bibinfo {year} {1994})},\ \Eprint
  {http://arxiv.org/abs/gr-qc/9402014} {arXiv:gr-qc/9402014 [gr-qc]}
  \BibitemShut {NoStop}%
\bibitem [{\citenamefont {Maggiore}(2007)}]{Maggiore:book}%
  \BibitemOpen
  \bibfield  {author} {\bibinfo {author} {\bibfnamefont {M.}~\bibnamefont
  {Maggiore}},\ }\href@noop {} {\emph {\bibinfo {title} {{Gravitational Waves.
  Vol. 1: Theory and Experiments}}}},\ Oxford Master Series in Physics\
  (\bibinfo  {publisher} {Oxford University Press},\ \bibinfo {year}
  {2007})\BibitemShut {NoStop}%
\bibitem [{\citenamefont {Poisson}\ and\ \citenamefont
  {Will}(1995)}]{Poisson:1995ef}%
  \BibitemOpen
  \bibfield  {author} {\bibinfo {author} {\bibfnamefont {E.}~\bibnamefont
  {Poisson}}\ and\ \bibinfo {author} {\bibfnamefont {C.~M.}\ \bibnamefont
  {Will}},\ }\href {\doibase 10.1103/PhysRevD.52.848} {\bibfield  {journal}
  {\bibinfo  {journal} {Phys. Rev. D}\ }\textbf {\bibinfo {volume} {52}},\
  \bibinfo {pages} {848} (\bibinfo {year} {1995})},\ \Eprint
  {http://arxiv.org/abs/gr-qc/9502040} {arXiv:gr-qc/9502040} \BibitemShut
  {NoStop}%
\bibitem [{\citenamefont {{Thrane}}\ \emph {et~al.}(2013)\citenamefont
  {{Thrane}}, \citenamefont {{Christensen}},\ and\ \citenamefont
  {{Schofield}}}]{2013PhRvD..87l3009T}%
  \BibitemOpen
  \bibfield  {author} {\bibinfo {author} {\bibfnamefont {E.}~\bibnamefont
  {{Thrane}}}, \bibinfo {author} {\bibfnamefont {N.}~\bibnamefont
  {{Christensen}}}, \ and\ \bibinfo {author} {\bibfnamefont {R.~M.~S.}\
  \bibnamefont {{Schofield}}},\ }\href {\doibase 10.1103/PhysRevD.87.123009}
  {\bibfield  {journal} {\bibinfo  {journal} {\prd}\ }\textbf {\bibinfo
  {volume} {87}},\ \bibinfo {eid} {123009} (\bibinfo {year} {2013})},\ \Eprint
  {http://arxiv.org/abs/1303.2613} {arXiv:1303.2613 [astro-ph.IM]} \BibitemShut
  {NoStop}%
\bibitem [{\citenamefont {{Thrane}}\ \emph {et~al.}(2014)\citenamefont
  {{Thrane}}, \citenamefont {{Christensen}}, \citenamefont {{Schofield}},\ and\
  \citenamefont {{Effler}}}]{2014PhRvD..90b3013T}%
  \BibitemOpen
  \bibfield  {author} {\bibinfo {author} {\bibfnamefont {E.}~\bibnamefont
  {{Thrane}}}, \bibinfo {author} {\bibfnamefont {N.}~\bibnamefont
  {{Christensen}}}, \bibinfo {author} {\bibfnamefont {R.~M.~S.}\ \bibnamefont
  {{Schofield}}}, \ and\ \bibinfo {author} {\bibfnamefont {A.}~\bibnamefont
  {{Effler}}},\ }\href {\doibase 10.1103/PhysRevD.90.023013} {\bibfield
  {journal} {\bibinfo  {journal} {Physical Review D}\ }\textbf {\bibinfo
  {volume} {90}},\ \bibinfo {eid} {023013} (\bibinfo {year} {2014})},\ \Eprint
  {http://arxiv.org/abs/1406.2367} {arXiv:1406.2367 [astro-ph.IM]} \BibitemShut
  {NoStop}%
\bibitem [{\citenamefont {{Himemoto}}\ and\ \citenamefont
  {{Taruya}}(2017)}]{2017PhRvD..96b2004H}%
  \BibitemOpen
  \bibfield  {author} {\bibinfo {author} {\bibfnamefont {Y.}~\bibnamefont
  {{Himemoto}}}\ and\ \bibinfo {author} {\bibfnamefont {A.}~\bibnamefont
  {{Taruya}}},\ }\href {\doibase 10.1103/PhysRevD.96.022004} {\bibfield
  {journal} {\bibinfo  {journal} {\prd}\ }\textbf {\bibinfo {volume} {96}},\
  \bibinfo {eid} {022004} (\bibinfo {year} {2017})},\ \Eprint
  {http://arxiv.org/abs/1704.07084} {arXiv:1704.07084 [astro-ph.IM]}
  \BibitemShut {NoStop}%
\bibitem [{\citenamefont {{Himemoto}}\ and\ \citenamefont
  {{Taruya}}(2019)}]{2019PhRvD.100h2001H}%
  \BibitemOpen
  \bibfield  {author} {\bibinfo {author} {\bibfnamefont {Y.}~\bibnamefont
  {{Himemoto}}}\ and\ \bibinfo {author} {\bibfnamefont {A.}~\bibnamefont
  {{Taruya}}},\ }\href {\doibase 10.1103/PhysRevD.100.082001} {\bibfield
  {journal} {\bibinfo  {journal} {\prd}\ }\textbf {\bibinfo {volume} {100}},\
  \bibinfo {eid} {082001} (\bibinfo {year} {2019})},\ \Eprint
  {http://arxiv.org/abs/1908.10635} {arXiv:1908.10635 [astro-ph.IM]}
  \BibitemShut {NoStop}%
\bibitem [{\citenamefont {{Meyers}}\ \emph {et~al.}(2020)\citenamefont
  {{Meyers}}, \citenamefont {{Martinovic}}, \citenamefont {{Christensen}},\
  and\ \citenamefont {{Sakellariadou}}}]{2020PhRvD.102j2005M}%
  \BibitemOpen
  \bibfield  {author} {\bibinfo {author} {\bibfnamefont {P.~M.}\ \bibnamefont
  {{Meyers}}}, \bibinfo {author} {\bibfnamefont {K.}~\bibnamefont
  {{Martinovic}}}, \bibinfo {author} {\bibfnamefont {N.}~\bibnamefont
  {{Christensen}}}, \ and\ \bibinfo {author} {\bibfnamefont {M.}~\bibnamefont
  {{Sakellariadou}}},\ }\href {\doibase 10.1103/PhysRevD.102.102005} {\bibfield
   {journal} {\bibinfo  {journal} {\prd}\ }\textbf {\bibinfo {volume} {102}},\
  \bibinfo {eid} {102005} (\bibinfo {year} {2020})},\ \Eprint
  {http://arxiv.org/abs/2008.00789} {arXiv:2008.00789 [gr-qc]} \BibitemShut
  {NoStop}%
\end{thebibliography}
\bibliographystyle{apsrev4-1}
%


\end{document}